\documentclass[acmsmall, authorversion]{acmart}

\usepackage{tabularx}
\usepackage{color,soul}
\usepackage{multirow}
\usepackage{tabularx}

\setcopyright{acmcopyright}
\copyrightyear{2023}
\acmYear{2023}
\acmDOI{10.1145/xxxxxx}

\acmConference[CSCW '23]{CSCW '23}{October 13--18, 2023}{Minneapolis, MN, USA}
\acmBooktitle{CSCW '23, October 13--18, Minneapolis, MN, USA}
\acmPrice{15.00}
\acmISBN{978-1-4503-XXXX-X}

\begin{document}
\title[The Effect of Explanations on Perceptions of Voice Assistants]{Ignorance is Bliss? The Effect of Explanations on Perceptions of Voice Assistants}

\author{William Seymour}
\orcid{0000-0002-0256-674}
\affiliation{
  \institution{King's College London}
  \streetaddress{Bush House, 30 Aldwych}
  \city{London}
  \country{UK}
  \postcode{WC2B 4BG}
}
\email{william.1.seymour@kcl.ac.uk}

\author{Jose Such}
\orcid{0000-0002-6041-178X}
\affiliation{
  \institution{King's College London}
  \streetaddress{Bush House, 30 Aldwych}
  \city{London}
  \country{UK}
  \postcode{WC2B 4BG}
}
\email{jose.such@kcl.ac.uk}

\renewcommand{\shortauthors}{Seymour and Such}

\begin{abstract}
Voice assistants offer a convenient and hands-free way of accessing computing in the home, but a key problem with speech as an interaction modality is how to scaffold accurate mental models of voice assistants, a task complicated by privacy and security concerns. We present the results of a survey of voice assistant users (n=1314) measuring trust, security, and privacy perceptions of voice assistants with varying levels of online functionality explained in different ways. We then asked participants to re-explain how these voice assistants worked, showing that while privacy explanations relieved privacy concerns, trust concerns were exacerbated by trust explanations. Participants' trust, privacy, and security perceptions also distinguished between first party online functionality from the voice assistant vendor and third party online functionality from other developers, and trust in vendors appeared to operate independently from device explanations. Our findings point to the use of analogies to guide users, targeting trust and privacy concerns, key improvements required from manufacturers, and implications for competition in the sector.
\end{abstract}


\maketitle

\section{Introduction}\label{sec:intro}
Domestic voice assistants such as Alexa and Google Assistant continue to grow in popularity, offering easy access to information and automation in the home. But the use of speech as the primary mode of interaction for these AI-driven devices obscures their underlying configuration and mechanics in ways that might otherwise be afforded by the design of their interfaces (e.g. the presence of a setting in a menu affords its changing, but this visibility does not easily translate to a conversation). Human-computer speech is presently a lower bandwidth interaction modality, as devices lack the ability to vary volume, intonation, and cadence. This leads to incomplete mental models of how they work~\cite{abdi2019more} and, combined with the always on and connected nature of voice assistants, has led to widespread privacy and security concerns about the extent to which they record and monetise what goes on around them~\cite{10.1145/3357236.3395501, 10.1145/3274371, 10.1145/3170427.3188448, abdi2019more}.

Recent work on people's perceptions of voice assistants reveals uncertainty about how voice assistants work, such as when devices are recording conversations, and what audio and transcripts are used for beyond the immediate actioning of requests~\cite{10.1145/3357236.3395501, malkin2018can, 10.1145/3274371}. Given that several major voice assistant manufacturers are global leaders in data-driven sectors such as advertising and e-commerce, the potential for conflicts of interest between profit and people's privacy are clear, and are reflective of wider concerns over the rise of surveillance capitalism~\cite{cannizzaro2020trust, zeng2017end}. There is an understanding that the main user-facing outcome of regulatory responses---privacy policies---inadequately address user concerns; often difficult to understand~\cite{10.1145/985692.985752}, contradictory~\cite{andow2019policylint}, or simply left blank~\cite{edu2021skillvet}, privacy policies have become something that the majority of users blindly click through without reading. 

The focus on `privacy by design' in the EU General Data Protection Regulation (GDPR, Art. 25) is a welcome development, and is leading to a consistent set of building blocks to engineer GDPR principles into systems~\cite{gurses2016privacy}. In particular, a number of design patterns and strategies are emerging towards this end~\cite{danezis2015privacy}, including those attempting to adequately inform users, provide actionable controls, and demonstrate compliance with stated policy goals~\cite{such2017privacy}. However, some can be difficult to operationalise effectively when applied to complex intelligent systems like voice assistants~\cite{such2017privacy} and this is further complicated by the highly inaccurate mental models that users currently have of them~\cite{abdi2019more}. 

Related to uncertainties about how voice assistants work are concerns over the reliability of speech recognition. Studies of voice assistant logs suggest that a significant amount of voice assistant activations fail in some way. \citeauthor{10.1145/3196709.3196772} identify 26\% of logged Alexa invocations as rapidly following another request, a characteristic feature of functional task failure, and 9.6\% of remaining activations as unprocessable or misfires~\cite{10.1145/3196709.3196772}. \citeauthor{dambanemuya2020alexa} similarly report a 30\% rate of irrelevant or failed responses to news-related requests~\cite{dambanemuya2020alexa}. This makes it easy to believe media reports of voice assistants recording conversations and sending them to others\footnote{https://www.theguardian.com/technology/2018/may/24/amazon-alexa-recorded-conversation}, and leads to general misgivings about the maturity of voice recognition technology~\cite{10.1145/3313831.3376529}. Examples of failed interactions in the literature suggest that an improved understanding of the underlying technology would help users recover from failed interactions~\cite{10.1145/3173574.3174214}. Given this, finding ways to improve mental models of voice assistants and other smart home technology represents a major open challenge for the HCI community.

As such, the paper addresses the following research questions:
\begin{enumerate}
    \item[RQ1] How does the presence of first and third party entities in a voice assistant affect users' trust, privacy, and security perceptions of these devices?
    \item[RQ2] How does the provision of trust, privacy, and security information given in an explanation of a voice assistant affect users' trust, privacy, and security perceptions?
\end{enumerate}

And in so doing makes the following contributions:
\begin{itemize}
    \item Shows that voice assistants with first and third party entities are perceived as more concerning than offline assistants, with finer distinctions often made between the inclusion of first and third party entities.
    \item Shows that voice assistant explanations involving information about trust increase trust concerns, while those involving information about privacy reduce privacy concerns.
    \item Demonstrates that for some aspects of trust, users' existing choice of device is more important than functionality or explanation in determining concerns about new devices
    \item Finds that the `incident anxiety' trust factor previously observed in other smart home devices also manifests in voice assistants.
\end{itemize}

More specifically, the paper presents the results of a study that sought to understand the effect of explanations on users' perceptions of voice assistants. We used a two level study design that described voice assistants with differing types of online connectivity using combinations of trust, privacy, and security information. We then tested to see how these factors impacted people's perceptions of the assistants and asked participants to explain how they worked back to us in their own words to see which words and concepts were retained. The results of the study reveal nuances around the introduction of third parties, suggesting security reservations around the use of third party skills over first party or local ones, privacy reservations around the use of any non-local skills, and different responses to different kinds of information. Privacy explanations were found to ameliorate privacy concerns, whereas trust explanations increased concerns over trustworthiness. We also identify functionality and risks that were both well and poorly understood by participants, such wake words and skill squatting respectively. Reflecting on this, we lay out recommendations around the targeting of different concerns through explanations, development of analogies to build mental models, and the unique relationship of trust and voice assistants.

\section{Background}
\subsection{Explainable Smart Technology and AI}
When considering explanations of intelligent systems such as voice assistants, an obvious related area of research is the burgeoning body of work on fairness, accountability, and transparency issues with ML-driven classifiers and decision-making systems. Explanations have been identified as a key tool in making these systems more understandable, promoting their safe, ethical, and effective use.

General work by ~\citeauthor{hoffman2018metrics} on the criteria for effective explanations suggests that they should (1) give a satisfying, detailed, and complete understanding of how the system works; (2) be actionable; and (3) indicate how reliable and trustworthy the system is~\cite{hoffman2018metrics}. There is a growing body of work on producing explanations for the actions of intelligent systems, addressing a wide range of problems communicating things such as the link between data and conclusions and uncertainty in those conclusions~\cite{mittelstadt2016ethics}, or the reasons and justifications for a recommendation~\cite{mosca2021elvira,mosca2022explainable}. Often used in the context of machine learning systems and decision aids (such as credit scoring algorithms), they are frequently driven by ethics and seek to both address information asymmetries between data processors and subjects, and serve as a human-readable check when decisions need to be audited. Common methods for achieving this include the approximation of local boundaries~\cite{ribeiro2016should} and generation of counterfactuals~\cite{wachter2017counterfactual}. 

When we consider the applicability of this work to the research questions of the present study several differences from the general explainable AI literature become clear. Whereas the goal of the latter is mostly to contextualise the output of a model to an expert or provide insights to end users (e.g. about what would have been required for a different decisions to be reached), voice assistant explanations as we envisage them in this paper are about providing mental scaffolding to help people develop accurate mental models of how their devices work. Unlike with automated systems offering bank loans, a major problem in everyday connected life is the \textit{lack} of visibility of algorithms (or `AI') in contemporary platforms, apps, and services. Even when people are aware of the learning processes that shape their interactions, understandings of how these systems work and the associated benefits and drawbacks is highly variable (e.g. as seen with social feeds~\cite{10.1145/3025453.3025659, eslami2016first, pybus2021did}). Work on visualising actions (including data sharing) by autonomous vehicles~\cite{koo2015did}, mobile apps~\cite{10.1145/3173574.3173967, 10.1145/2556288.2557421} and smart home devices~\cite{10.1145/3313831.3376264} has been developing strategies for further understanding how people intuitively interpret the information available to them, as well as developing techniques for explaining why and how devices behave as they do. These include the use of data to support personal strategies, information about \emph{why} actions were taken, and the business models that drive observed behaviour. These efforts have been enabled by technical endeavours to reverse-engineer network flows from devices~\cite{10.1145/3397333, 10.1145/3395351.3399421}, as for many devices exact information about where and how frequently information is shared is unavailable.

Many of these solutions deliver some form of privacy notice, and work on the design of these notices adopts similar recommendations to \citeauthor{hoffman2018metrics}, suggesting that they be relevant, actionable, and understandable. Specific recommendations that notices (1) understand the system’s data practices and users; (2) are short and specific; (3) highlight unexpected practices; and (4) provide details on demand~\cite{schaub2017designing}, provide insights that are potentially applicable to voice assistant explanations. Given that unactionable privacy revelations often leave users feeling trapped and helpless~\cite{10.1145/3313831.3376529, 10.1145/2556288.2557421}, this looks to be a key part of any implementation. Attempts to adapt the above to actual smart devices usually follow a design-led approach; the fictional Polly smart kettle displays data flows on the side of the kettle, and presents an explicit trade-off between functionality and data collection~\cite{lindley2017internet}, and other approaches have focused on representing ambiguity around data collection and sharing~\cite{coulton2019more} or the wider smart home design space~\cite{chiang2020exploring}. Unfortunately this often runs counter to the prevailing minimalist design language that prioritises sleek and unobtrusive form factors over those that have more affordances and relay a richer set of information (summed up by \citeauthor{10.1145/3173574.3174123} as ``appiness''~\cite{10.1145/3173574.3174123}), and this phenomena is only exacerbated by the hands-free nature of voice assistants that further rules out visual cues as a means of reliably communicating information about a device.

\subsection{Voice Assistant Privacy and Security: Perceptions and Realities}
Concerns over privacy and security are frequently found amongst users of voice assistants. It is not uncommon, for example, to find folk theories describing hacking as a bigger threat to privacy than data collection through surveillance capitalism~\cite{zimmermann2019assessing} or conversations in the vicinity of VAs being used to target advertising~\cite{10.1145/3357236.3395501} (an activity perceived much more negatively than data collection by companies for other purposes~\cite{10.1145/3375188}). The knowledge that one is being recorded can lead to negative consequences, with prior work showing the damaging long term psychological effects of ubiquitous surveillance in the home~\cite{10.1145/2370216.2370224}.

But perceptions of privacy and security issues are often complicated by incomplete mental models of voice assistants, particularly when devices occupy spaces shared with cohabitant users~\cite{10.1145/3313831.3376529} or non-user `bystanders'~\cite{10.1145/3419249.3420164, 10.1145/3359161, 10.1145/3313831.3376529}. This lack of understanding leads to a general lack of awareness about the risks involved~\cite{abdi2019more, 10.1145/3274371}, leading to people falling back on non-technical coping strategies in an attempt to protect themselves~\cite{abdi2019more, 10.1145/3313831.3376264}. This is not helped by the fact that research has also overly focused the risks of human-voice assistant interaction at the expense of other parts of the voice assistant ecosystem~\cite{10.1145/3412383}, in turn skewing media coverage of privacy and security risks to users' everyday interactions.

Research on voice assistant `skills' or `actions', which extend the capabilities of voice assistants via conversational software developed by third parties (e.g. a Spotify skill to play music from a Spotify account), and the ecosystems they form highlights security and privacy problems with skill vetting, permissions, `squatting', and data handling ~\cite{edu2021skillvet, kumar2018skill, lentzsch2021hey, edu2022measuring}. More generally, people without technical backgrounds are less likely to understand the organisations and infrastructure that their devices are integrated with, including the risks to privacy and security that arise as a result~\cite{kang2015my}. Prior work also shows that knowledge of where data is being stored, processed, and shared by assistants is at best localised to users' particular brand of assistant~\cite{abdi2019more}. This lack of knowledge leads to the avoidance of features that are perceived to be more risky, such as online shopping using a voice assistant~\cite{abdi2019more, 10.1145/3274371}.

The small amount of systems security research that does exist in this area also demonstrates the dangers involved when connecting voice assistants to security-critical devices such as locks and garage doors; proof-of-concept attacks that utilise lasers~\cite{sugawara2020light} or high frequency sound~\cite{10.1145/3133956.3134052} to manipulate voice assistant devices. These attacks are not typically seen in the wild, but further give the impression that voice assistants were not designed to provide the level of security required for the tasks they can be asked to do. Several versions of privacy and security labels have been proposed in order to communicate potential risks to consumers and allow for the comparison of factors such as availability of security updates between devices~\cite{10.1145/1572532.1572538, 10.1145/3290605.3300764}. There has also been legal debate over the way that voice assistants often store audio recordings of users in the cloud where they can be accessed by law enforcement, bypassing existing protections on activities undertaken in the home (such as the Third Amendment in the US)~\cite{dunin2020alexa}.

\subsection{Trust and Voice Assistants as Social Actors}~\label{sec:interactions}
Unlike traditional visual interfaces where there are established design patterns for communicating options and error states, the design language of the speech interfaces used in voice assistants is less mature~\cite{10.1145/3313831.3376522}, with ongoing research into fundamental aspects such as accessibility~\cite{10.1145/3313831.3376225}, progresivity~\cite{10.1145/3342775.3342788}, and conversational repair~\cite{10.1145/3392838}. Alongside routine mis-interpretation and accidental activations, outbreaks of creepy laughter~\cite{lee2018amazon} and wake words being triggered by advertisements~\cite{burgess2017google} have achieved high media visibility and furthered perceptions that users cannot trust these devices to work consistently and that they are potentially unsettling.

Gathering user perceptions about the reliability of voice assistants poses an interesting problem. On the one hand, conventional measures of reliability in AI/intelligent systems (e.g.~\cite{madsen2000measuring}, closely related to trust~\cite{10.1145/3313831.3376551}) can be used to measure beliefs around task fulfilment, consistency over repeated interactions, and response accuracy. On the other hand, some studies have also highlighted the tendency for people to have satisfactory interactions with voice assistants even when they fail to complete the tasks given to them~\cite{lopatovska2018talk}, possibly related to situations where social interactions \textit{are} the main intended outcome of an interaction (as in ~\cite{10.1145/3359316}).

This is part of a larger tendency of people to anthropomorphise computers, particularly those that use speech. Early work by \citeauthor{10.1145/191666.191703} showed that people subconsciously treat computers as social actors, and do this whilst being fully aware that they are interacting computers rather than with people~\cite{10.1145/191666.191703}. At the same time, voice assistants are often specifically \textit{designed} to be anthropomorphised by users, leading researchers to investigate the extent to which they are (often implicitly) gendered and personified~\cite{abercrombie2021alexa, 10.1145/3359316}. Prior work has shown that the use of these mechanisms is associated with increased information disclosure, mindless enactment of social scripts, and more positive responses to information sharing~\cite{10.1086/209566, 10.1145/3176349.3176868, 10.1145/3375188}.

Relatedly, work on partner models in speech interfaces shows the significant role they play in interaction. These models are generally composed of perceptions of a system's competence and dependability, it's human-likeness, and cognitive flexibility~\cite{10.1145/3411764.3445206}. While people generally assume greater knowledge when interacting with voice interfaces than human interlocutors, they use assumptions about what other \textit{people} know when estimating the knowledge of artificial partners~\cite{cowan2017they}. It is worth noting, however, that greater human likeness in VAs also increases expectations of performance and can therefore reinforce the limitations of current solutions.\cite{10.1145/3098279.3098539}

These phenomena each add an additional layer of complexity to how people conceptualise voice assistants, blending rational and social ways of thinking; drawing studies of both adult and younger voice assistant users show a wide variety of depictions including humans, machines, and extraterrestrial satellites~\cite{10.1145/3290607.3312796, 10.1145/3313831.3376416}. These variations hint at deeper differences in people's mental models, such as the extent to which assistants are perceived to be shared between or unique to different users (e.g. whether there is one communal Alexa, or millions of individual instantiations)~\cite{10.1145/3290607.3312796}.

\subsection{Designing the Voice Assistants of the Future}
Looking to the future, where might voice assistant explanations fit in relation to other developments? Works envisaging voice assistants as virtual butlers~\cite{payr2013virtual}, councillors~\cite{kannampallil2020cognitive}, and music  coaches~\cite{10.1145/3313831.3376493}, or as part of smart homes imbued with different `personalities'~\cite{10.1145/2971648.2971757} present intriguing variations to the social mechanisms typically seen with voice assistants used in the home. Developing techniques in verification and monitoring for ML systems~\cite{batten2021efficient, huang2017safety, criado2016selective} also make it increasingly possible that future generations of voice assistants and smart devices could have provable privacy and security properties like those seen in cryptographic algorithms or mission-critical software (e.g. the Signal protocol~\cite{cohn2020formal} and the SEL4 microkernel~\cite{klein2009sel4}). We believe that explanations are well positioned to ease both of these transitions, for the same reasons that they are applicable to the assistants of today. Easing the transition towards assistants that are capable of being more proactive and engaging users in more of a discussion will require the ability develop trust in them and an understanding of how and why they might take certain actions.

\subsection{Summary}
In seeking to better understand our interactions with voice assistants and adjacent smart devices, researchers have uncovered a number of factors that contribute to a widespread lack of understanding about many aspects of these devices, particularly concerning privacy, security, and trust. These often involve misconceptions around the technical capabilities of devices, such as when they are recording, that are exacerbated by the `appiness'~\cite{10.1145/3173574.3174123} of sleek but deliberately low-bandwidth interfaces and the more socially-oriented cognitive processes that are engaged when interacting with devices via speech.

Current answers to these problems revolve around visual means---utilising visualisations and printed labels---which are less suitable for use with voice assistants. Prior work on explanations in adjacent contexts is extremely promising, but fails to address a number of phenomena that are specific to voice assistants, such as misfires and anthropomorphism. Therefore, we now explore responses to a number of different vignettes that describe functionality specific to voice assistants and could feasibly be delivered piecemeal via speech. By studying responses to these descriptions, and how well people are able to explain them in their own words, we hope to discover which approaches are the most effective and the effect that they have on people’s perceptions of voice assistant technology.

\section{Methods}
In order to explore the research questions above we constructed a survey of voice assistant users exploring perceptions of the voice assistant ecosystem and responses to explanations. The survey consisted of three main parts: the first presented users with a hypothetical voice assistant, an explanation of how it worked, and assessed their perceptions of its trustworthiness, privacy, and security; the second asked participants to re-explain the workings of this assistant in their own words; and the final part of the survey gathered baseline levels of privacy and security concern. Following \citeauthor{pavlou2006understanding} we adopted a definition of trust as ``the belief that an entity will act cooperatively to fulfil clients’ expectations without exploiting their vulnerabilities''~\cite{pavlou2006understanding}. This includes trust in devices, vendors, and other entities that are explained in more detail below. The survey was implemented using Qualtrics and we recruited via the Prolific Academic platform\footnote{\url{https://prolific.co}}, compensating participants at an average rate of \pounds8.70 an hour. Participants were all 18+ years of age, resident in the UK, and owned a voice assistant. All parts of the study were approved by our institution's IRB. Survey questions, anonymised results, and the qualitative codebook are provided as supplemental material and archived at \url{https://osf.io/z7f5n}.

\subsection{Designing the Vignette Explanations}
To investigate changes in perceptions towards assistants with different functionality explained in different ways, we utilised a two level (3x8) study design visualised in Table~\ref{tab:studydesign}. We generated a set of 24 vignettes describing three fictional voice assistants with differing numbers of external entities, each explained in eight different ways. Entities were represented through the skills available to the assistant: offline/ on-device skills only (0E), first party online skills (1E), or third party online skills (3E). Explanations were generated from content snippets containing: functional descriptions of how a voice assistant would respond to a command (F); information about whether users could trust the assistant to function reliably and consistently, including ways that it might violate the interaction context~\cite{schaub2017designing} (T); information about data privacy (P); and information about the security of the device (S). All vignettes included a basic functional description of the assistant, with every combination of trust, privacy, and security represented for a total of 8 explanation combinations (hereafter F denotes an explanation containing \textit{only} the functional description). We refer to these vignettes in the remainder of the paper using the shorthand given above in brackets, e.g. the vignette for third party entities with functional, trust, privacy, and security information would be 3E/TPS. An annotated example of this vignette is given Figure~\ref{fig:vig3}.

Conversational design for contemporary voice assistants commonly employs a model of \textit{intents}, \textit{utterances}, and \textit{slots}. An intent represents an action that the user can perform (e.g. playing music), which can contain variables called slots (e.g. an artist or song title), and are invoked through utterances (e.g. ``Play something by Taylor Swift''). We represented this approach in the vignettes by saying that the assistant software matched key words in a request to determine both the command to run and any variables stating how it should be executed. In order to minimise the potential for priming participants towards anthropomorphism we gave the voice assistant an invocation that was not a human name (``hey assistant''), and did not directly include words said by the assistant in the vignette. Prior work has suggested that the use of data-based framings of smart home devices can encourage creativity when imagining uses for data and devices~\cite{clark2017devices}, so we were specific about the kinds of information and processing that were taking place through the assistant.

\begin{table}
    \centering
    \begin{tabularx}{385pt}{|X|X|X|}
    \toprule
    \multicolumn{3}{|c|}{An assistant with the following connectivity:} \\ \hline
    Offline functionality only (0E) & Online functionality with the first-party vendor/entity (1E) & Online functionality with third party entities (3E) \\ \hline
    \multicolumn{3}{|c|}{Explained using one of the following combinations of details:} \\ \hline
    Functionality (F) & Functionality \& Trust (T) & Functionality \& Privacy (P) \\ \hline 
    Functionality \& Security (S) & Functionality, Trust \& Privacy (TP) & Functionality, Trust \& Security (TS) \\ \hline
    Functionality, Privacy \& Security (PS) & Functionality, Trust, Privacy \& Security (TPS) & - \\
    \bottomrule
    \end{tabularx}
    \caption{Overview of the 3x8 study design used to generate vignettes.}
    \label{tab:studydesign}
\end{table}

\begin{figure}
\begin{tabularx}{\textwidth}{l|X}
Functionality & Morgan uses the assistant through a smart speaker---``Hey Assistant, ask Pact Coffee to order me another bag of coffee''. The device keeps a recording of the last five seconds of what was said and checks to see if this contains the words ``Hey Assistant''. \\
Trust & Sometimes other things Morgan says sounds similar to this. This can lead to the assistant accidentally recording conversations and trying to interpret them as commands. \\
Security & The assistant can be configured to tell people apart based on their voices, but Morgan hasn’t set this up yet. Sometimes Morgan’s son tries to buy things with it. \\
&\\
Privacy & The recording of Morgan saying the request is sent over the internet to the assistant’s provider and are stored on the provider’s servers and are used to personalise skills and improve speech recognition algorithms.  \\
Trust & The assistant will stop working if it can't connect to the Internet. \\
Functionality & An algorithm on the developer’s server creates a transcript of Morgan’s speech, and tries to match key words to a list of possible commands. Once it matches the words order and Pact Coffee, it forwards the request on to the store where Morgan normally buys coffee along with details about where Morgan lives so that the coffee can be delivered, and their card details. This gets arranged into sentences and sent back to the device. It is then turned into speech using Morgan's preferred voice and read out loud. \\
Privacy & A copy of the interaction is saved in the voice assistant’s log. \\
\\
Trust & Sometimes the assistant mishears and opens the wrong skill, or accidentally orders something \\
Privacy & causing Morgan’s address and card details to be sent to the wrong company. \\
Functionality & The coffee company uses Morgan’s address and card details to generate the order and sends confirmation of this back to the assistant provider. \\
Privacy & Each company has a different policy about what data it collects from Morgan and how long it is stored for. \\
Security & Some skills might deliberately be named to share an invocation with another skill (I.e. they are both called `BBC'), or attempt to collect personal information without asking for the proper permissions. This can be used for fraud or to compromise Amazon or other online accounts. \\
\end{tabularx}
    \caption{Treatment 3E/TPS: voice assistant with first and third party skills, explanation includes functionality, trust, privacy, and security elements. Annotated to show which elements of the explanation correspond with the four information types.}
    \label{fig:vig3}
\end{figure}

\subsection{Measuring Trust Perceptions}
After presenting participants with a vignette we asked about their perceptions of its trustworthiness, including trust in its privacy and security, using eight questions from \citeauthor{cannizzaro2020trust} (originally Q8--Q15, reported here as Q1--Q8). The list of questions is given in Figure~\ref{fig:inventory}. While these all pertained to trust, they were divided into general trust in the device and its manufacturer (expressed as a combination of competence/performing tasks reliably, benevolence of vendors in respecting users' interests, and vendor's integrity~\cite{mayer1995integrative}), trust in the privacy of the device, and trust in the security of the device. To aid readability we refer to these three dimensions as trust, privacy, and security in the remainder of the paper. These questions were answered using a seven point Likert scale from strongly agree to strongly disagree, with `smart home devices' replaced with `voice assistant'. While both the privacy and security questions make use of the word security, the former describe outcomes affecting only data and the latter ones where this leads to follow-on impacts such as fraud (users were not given the question classifications). The same distinction applies to the privacy and security components of the vignette explanations. While no validated questionnaire exists for these concepts in this context, the questions were determined to have content (logical) validity by the original authors~\cite{cannizzaro2020trust}. In order to remove unnecessary variables we did not include information about the vendor of the fictional voice assistant in the vignettes and the trust component of the vignettes therefore mainly related to trust as competence, a limitation described further in Section~\ref{sec:limitations}.

\begin{figure}
\begin{enumerate}
    \item I would fully trust the voice assistant not to fail, and to function as I expect it to (general trust, reverse coded)
    \item Knowing that the voice assistant allowed companies or organisations to collect data about how I used it, and hence about my domestic habits, would restrict me from owning/using it (general trust)
    \item I would trust the manufacturer not to use data produced by the voice assistant for any purpose without my explicit consent (general trust, reverse coded)
    \item I think the likelihood of the security of voice assistants being compromised and resulting in a privacy/data breach is high (trust in privacy) 
    \item I think the impact of the security of voice assistants being compromised and resulting in a privacy/data breach is low (trust in privacy, reverse coded) 
    \item The Facebook user data sharing controversy makes me less willing to own/ use voice assistants (privacy) 
    \item I think the likelihood of the security of voice assistants being compromised and resulting in an incident (e.g. burglary, fraud) is high (trust in security) 
    \item I think the impact of the security of voice assistants being compromised and resulting in an incident (e.g. burglary, fraud) is low (trust in security, reverse coded) 
\end{enumerate}
\caption{Survey questions adapted from \citeauthor{cannizzaro2020trust}~\cite{cannizzaro2020trust}.}
\label{fig:inventory}
\end{figure}

\subsection{User-Written Explanations}
We also wanted to understand how easily participants were able to absorb and retain the concepts described in the different explanations. Participants were asked to re-explain how the assistant worked in their own words, as if they were describing it to a friend or family member (minimum of 150 characters/approx. 35 words). In order to balance the likelihood of participants copying the vignette explanation with the burden of unexpectedly being asked to remember it, we asked them to give their answer on the next page of the survey but allowed them to go back and re-read the explanation if they wished.

\subsection{Gathering Baseline Perceptions and Attitudes}
To avoid priming the survey closed by asking participants whether they felt that data shared with the assistant was more private or secure than that stored on their own personal computer, followed by standard questions about participants' general privacy and security practices and attitudes (an approach taken in prior work, such as~\cite{naeini2017privacy}). To assess privacy attitudes we used the 10 questions from the core of the Internet Users’ Information Privacy Concerns inventory (IUIPC)~\cite{malhotra2004internet} that measure attitudes towards control, awareness, and collection in the context of online privacy. For security attitudes we used all six questions from the SA-6 inventory of security attitudes~\cite{faklaris2019self}. IUIPC was developed and validated based on data from a total of 742 interviews, and SA-6 was developed and validated based on a total of 687 survey responses.

\subsection{Analysis}
\label{subsec:analysis}
Before opening the survey, a priori power analysis revealed that the minimum number of participants required to detect a `small' effect of size 0.2 across 3x8 groups at the $\alpha=0.05$ level would be 693 (29 per group). We therefore ensured that the number of collected responses would exceed this level.

Calculations were carried out using Python, and a record of the pre-processing and statistical tests used is provided in the form of a Python notebook in the supplementary materials and OSF repository. 
Given the presence of an underlying factor in the smart home survey that originally used Q1--8 from \citeauthor{cannizzaro2020trust} (`incident anxiety'), we were interested in testing for the presence of similar factor(s). This would indicate correlation between question responses that could then be taken into account. The suitability of the data for exploratory factor analysis (EFA) was verified via Bartlett’s Test of Sphericity and the Kaiser-Meyer-Olkin Test (KMO). The significance of the first (3369.53, p=0.0) indicates that the data has a sufficient amount of intercorrelation for factor analysis, and the ``meritorious''~\cite{kaiser1974little} KMO value (0.84) that the results may be useful in understanding the data via factor analysis.

We used ANOVA tests to identify the presence of significant effects on responses to the eight inventory questions followed with Tukey HSD tests, the latter of which corrects for Type I errors associated with testing multiple hypotheses simultaneously. These were used to identify the specific pairs of functionality or explanation levels that had significant differences in response means. Answers to baseline perception questions on privacy and security were summed to give an overall score for comparison. Pearson's $\rho$ was used to calculate linear correlations. Answers to questions using a Likert scale were coded from 1 to 7 such that greater scores indicate greater concern, with questions reverse coded as appropriate. Mean differences between experiment conditions are reported using arrow notation to show the groups and direction of increase (e.g. $\Delta\bar{x}$ of 0.5 for 0E $\rightarrow$ 1E means that the average mean for 1E was 0.5 higher than for 0E).

For the participant-provided re-explanations, a sample of 400 responses was chosen to achieve good coverage of the data given the shorter nature of the responses, and contained a balance of the 24 different study conditions. One researcher analysed 200 of these responses to create a thematic codebook following Braun and Clarke~\cite{braun2012thematic}. The codebook was independently applied to the same sample by a second researcher, with a calculated agreement of $\kappa=0.86$. While there is no agreed upon threshold for sufficiency when interpreting Cohen's Kappa, this value is large enough to suggest a reasonably descriptive and objective codebook~\cite{landis1977measurement, fleiss2013statistical}. The remaining 200 responses in the sample were then divided between the researchers and coded using the codebook. Of particular interest during the qualitative analysis were the words and concepts used by participants to describe their understanding of voice assistants, drawing out the ideas that were most commonly internalised and the analogies used to understand this complex technology.

\section{Results}
\subsection{Demographics and Baseline Perceptions}
A total of 1314 survey responses were received, with 24 responses rejected through the Prolific platform and excluded from the analysis. Reasons for this included incomplete answers to the survey questions, revoked consent, empty re-explanations, and verbatim repetition of the vignette explanation in participants' re-explanations.

This left 1290 full responses for analysis. Participants were aged between 18 and 83 (mean=35.4, median=34, $\sigma$=12.0), with 49.8\% of participants identifying as women. A full breakdown of participant demographics is given in Table~\ref{tab:demographics}. As expected, baseline perceptions had an effect on the responses of participants to Q1--8. Participants that had higher general privacy concerns were more likely to show higher privacy and security concerns about the hypothetical assistants in Q4, 6, and 7 ($0.10 < \rho < 0.14$), but showed lower trust concerns in Q2 ($\rho = 0.13$). Greater baseline security attitudes resulted in greater concern about trust, privacy, and security for all questions except Q2, where there was a slight negative effect ($0.18 < \rho < 0.34$ and $\rho = -0.37$ respectively). In this respect, the focus of Q2 on the companies and organisations associated with the voice assistant may have changed the way that participants answered that particular question, tapping into negative perceptions of the major companies operating in this space. Age was weakly correlated with greater trust concerns in Q1 ($\rho = 0.07$), and on average women rated the likelihood and impact of privacy and security events as higher than men in Q4, 5, 7, and 8 ($0.26 < \Delta\bar{x} < 0.38$). Owners of devices using Siri were much more likely to have privacy and security concerns than those using Alexa or the Google Assistant for Q6 and Q7 ($0.39 < \Delta\bar{x} < 0.52$). On the other hand, Alexa users were more likely to express trust concerns in Q2 than those using Siri ($\Delta\bar{x} = 0.37$). All results reported above were significant at the $p >$ 0.05 level.

\begin{table}
    \centering
    \begin{tabular}{|l|l|rr|}
    \toprule
    \multirow{5}{5em}{Age} & 18--28 & 32.6\% & (421) \\
    & 29-38 & 30.8\% & (397) \\
    & 39-48 & 19.1\% & (247) \\
    & 49-58 & 11.4\% & (147) \\
    & 59+ & 4.6\% & (59) \\
    \hline
    \multirow{3}{5em}{Gender Id.} & Men & 50.1\% & (647) \\
    & Women & 49.8\% & (643) \\
    & Other/None & 0.1\% & (1) \\
    \hline
    \multirow{4}{5em}{Assistant} & Amazon Alexa & 55.7\% & (719) \\
    & Google Assistant & 26.3\% & (339) \\
    & Siri & 16.0\% & (207) \\
    & Other & 2.0\% & (26) \\
    \hline
    \multirow{3}{5em}{Computer Use at Work} & >$\frac{2}{3}$ of time & 59.3\% & (766) \\
    & $\frac{1}{3}$ to $\frac{2}{3}$ of time & 13.6\% & (176) \\
    & <$\frac{1}{3}$ of time & 13.2\% & (170) \\
    & None or N/A & 13.9\% & (179) \\
    \bottomrule
    \end{tabular}
    \caption{Demographic information for survey participants.}
    \label{tab:demographics}
\end{table}

\subsection{Exploratory Factor Analysis}
After the Bartlett and KMO tests described in the methodology suggested that some patterns in the survey data may be caused by underlying factor(s), we conducted exploratory factor analysis. Applying Velicer’s Minimum Average Partial method~\cite{velicer1976determining} to Questions 1--8 suggested the retention of a single factor with a proportional variance of 39\%. Comparing to ~\citeauthor{cannizzaro2020trust} we found a similar distribution of eigenvalues, with only one eigenvalue significantly above 1 (the original method used for factor retention). In order to simplify the resulting factor and to allow easy comparison with prior work we retained the three questions loading at or above 0.7 (Q4, Q7, and Q8) and computed a weighted factor score for each participant. Unlike in prior work, this factor was not found to significantly correlate with trust in competence: the variables present in the underlying factor are the same as \cite{cannizzaro2020trust} with the exception of Q1, which was not included based on analysis of the current data set. These questions are centred around the likelihood of an adverse event and resulting physical risk (`incident anxiety'), and the similarity of questions and context suggest that a closely related phenomena exists with VAs. Factor loadings are given in Table~\ref{tab:factor_loading}. The results of ANOVA tests showed that there were significant interactions between incident anxiety and explanation content ($p=0.0027$), online functionality ($p<0.001$), and the interaction between the two ($p<0.014$). Follow up Tukey HSD tests showed significant mean differences between all levels of online functionality, as well as between privacy explanations and those with security or security and reliability. These are included at the bottom of Table~\ref{tab:anova}. These findings suggest that the introduction of online functionality and the absence of privacy explanations both contribute to a general sense of incident anxiety in users.

\begin{table}
    \centering
    \begin{tabular}{r|r}
        \toprule
        Question & Factor Loading \\
        Q1 Trust in competence & 0.51 \\
        Q2 Trust in benevolence & -0.57 \\
        Q3 Trust in integrity & 0.51 \\
        \textbf{Q4 Likelihood of a privacy breach} & \textbf{0.75} \\
        Q5 Impact of a privacy breach & 0.62 \\
        Q6 Impact of controversy & 0.50 \\
        \textbf{Q7 Likelihood of incident} & \textbf{0.78} \\
        \textbf{Q8 Impact of incident} & \textbf{0.70} \\
        \bottomrule
    \end{tabular}
    \caption{Factor loadings for the eight trust questions, with items in bold retained for calculating participant factor scores. Note when comparing with \cite{cannizzaro2020trust} that values have been inverted such that positive values indicate greater concern.}
    \label{tab:factor_loading}
\end{table}

\subsection{Differences Between Explanation Groups}
In order to further explore the survey data we subsequently conducted a series of two way ANOVA tests on the functionality and explanation factors, which unsurprisingly revealed that both explained a significant amount of variance between the survey treatments for all questions in the inventory except Q2. There were no significant interactions between the effects of functionality and explanation for the individual questions except at a very small scale when considering the impact of security breaches (Q8, $\omega^2$=0.009, p=0.025). Follow-up Tukey tests revealed a number of significant differences between the means of individual functionality and explanation content treatments. These are shown in Table~\ref{tab:anova}. We briefly describe these results here to the extent that they extend and supplement the factor analysis.

\begin{table}
    \centering
    \begin{tabularx}{350pt}{|X|l|r|}
    \toprule
    Question & Conditions & Mean Difference \\
    \hline
    \multirow{10}{*}{Q1---Trust in competence (general trust)} & 0E $\rightarrow$ 1E & 0.50** \\
    & 0E $\rightarrow$ 3E & 0.51** \\
    \cline{2-3}
    & F $\rightarrow$ TS  & 0.53*$\;$ \\
    & P $\rightarrow$ T   & 0.81** \\
    & P $\rightarrow$ TP  & 0.84** \\
    & P $\rightarrow$ TPS & 0.67** \\
    & P $\rightarrow$ TS  & 0.85** \\
    & P $\rightarrow$ S   & 0.66** \\
    & PS $\rightarrow$ TP & 0.53*$\;$ \\
    & PS $\rightarrow$ TS & 0.54*$\;$ \\
    \hline
    \multirow{1}{*}{Q2---Trust in benevolence (general trust)} & None & - \\
    \hline
    \multirow{1}{*}{Q3---Trust in integrity (general trust)} & 0E $\rightarrow$ 3E & 0.33*$\;$ \\
    \hline
    \multirow{7}{*}{Q4---Likelihood of a privacy breach (privacy)} & 0E $\rightarrow$ 1E & 0.53** \\
    & 0E $\rightarrow$ 3E & 0.61** \\
    \cline{2-3}
    & P $\rightarrow$ T   & 0.66** \\
    & P $\rightarrow$ TS  & 0.69** \\
    & P $\rightarrow$ S   & 0.61** \\
    \hline
    \multirow{4}{*}{Q5---Impact of a privacy breach (privacy)} & 0E $\rightarrow$ 1E & 0.32*$\;$ \\
    & 0E $\rightarrow$ 3E & 0.48** \\
    \cline{2-3}
    & P $\rightarrow$ T   & 0.58*$\;$ \\
    & P $\rightarrow$ TS  & 0.53*$\;$ \\
    \hline
    \multirow{1}{*}{Q6---Impact of controversy (privacy)} & 0E $\rightarrow$ 3E & 0.33*$\;$ \\
    \hline
    \multirow{4}{*}{Q7---Likelihood of incident (security)} & 0E $\rightarrow$ 1E & 0.56** \\
    & 0E $\rightarrow$ 3E & 0.83** \\
    & 1E $\rightarrow$ 3E & 0.27*$\;$ \\
    \cline{2-3}
    & P $\rightarrow$ TS & 0.60*$\;$ \\
    \hline
    \multirow{3}{*}{Q8---Impact of incident (security)} & 0E $\rightarrow$ 1E & 0.53** \\
     & 0E $\rightarrow$ 3E & 0.82** \\
     & 1E $\rightarrow$ 3E & 0.29*$\;$ \\
     \hline
    \multirow{2}{*}{Q9---Privacy Relative to Own Device} & 0E $\rightarrow$ 3E & 0.36** \\
    & 1E $\rightarrow$ 3E & 0.28*$\;$ \\
    \hline
    \multirow{2}{*}{Q10---Security Relative to Own Device} & 0E $\rightarrow$ 3E & 0.39** \\
    & 1E $\rightarrow$ 3E & 0.27*$\;$ \\
    \cline{2-3}
    & P $\rightarrow$ TS & 0.50*$\;$ \\
    \hline
    \multirow{5}{*}{Underlying factor---``incident anxiety''} & 0E $\rightarrow$ 1E & 1.21** \\
    & 0E $\rightarrow$ 3E & 1.67** \\
    & 1E $\rightarrow$ 3E & 0.47*$\;$ \\
    \cline{2-3}
    & P $\rightarrow$ TS & 1.30** \\
    & P $\rightarrow$ S & 1.15*$\;$ \\
    \bottomrule
    \end{tabularx}
      \caption{Significant mean differences between survey treatments. Higher means indicate greater concern. A * indicates an adjusted $p \leq 0.05$ and ** an adjusted $p \leq 0.005$. Descriptions of the questions are taken from~\cite{cannizzaro2020trust}. Tests that did not meet the 0.05 significance threshold are not shown.}
    \label{tab:anova}
\end{table}

The introduction of external entities was unsurprisingly considered to come at the cost of trust, privacy, and security. Significant mean differences were observed between offline and online assistants (i.e. 0E $\rightarrow$ 1E or 3E) in nine of the ten questions, in line with previous literature (e.g.~\cite{zimmermann2019assessing}). What was more interesting were different ways in which this manifested: for security each additional level of external entities produced a significant increase in concerns (0E $\rightarrow$ 1E $\rightarrow$ 3E). For privacy the addition of any external entity produced an increase in concerns, with third party entities causing more than first party ones but not with enough of a distinction to show significant differences between first and third parties. This shows that people can and do distinguish between the number of external parties involved, with differing granularities of concern depending on what is at stake. This is true whether or not participants were aware of the existence of third party software on their own voice assistants or other devices, with the responses to Q10 largely suggesting the latter.

Vignette explanations that only contained privacy information were often associated with lower levels of concern than those that did not contain privacy information (i.e. the removal of privacy explanations often increased concerns about trust, privacy, \textit{and} security). Another key signal in the results was that for Q1 (which asked about trust in competence), the addition of trust information to almost any explanation without it led to significantly higher trust concerns. This effect was present even when both explanations contained privacy information, suggesting that it was a separate effect from the one described above.

The two questions gauging trust in benevolence and integrity (Q2 and Q3) were only associated with one significant effect between them (0E $\rightarrow$ 3E). Given the focus of these items on device vendors, this indicates that the extent to which participants trust a device's manufacturer is not affected by the way that its use is explained, and only slightly by the presence of online capability. Notably, by distinguishing only between 0E and 3E, participants seem to trust online first party services as much as they trust the device itself, with the addition of external entities (e.g. third-party skills providers) triggering a drop in trust.

\subsection{User-Written Re-explanations}\label{sec:explanations}
Most participants were able to explain the concept of using wake words to initiate interactions with the voice assistant. While this was often linked with the assistant beginning to `listen' for commands, a significant number of responses explicitly linked the activation/listening of the assistant with the \textit{recording} of speech for either wake word detection or the interpretation of the commands. The proportion of responses coded for wake words and recording did not significantly vary between connectivity levels or explanation content, except for vignette explanations covering all content types (\textit{RPS}) which were consistently higher.

The most common way that participants referred to the assistant was simply as `the assistant', followed by demonstrative pronouns (e.g. `it'), and personal pronouns (e.g. `she'). Referring to the assistant as if it were a machine, computer or robot was much less common. Some participants included a description of how the assistant matched key words in their requests to possible commands and variables, but this more mechanical way of thinking did not seem to bring with it a noticeable shift in the way that participants did (or did not) anthropomorphise the device. In re-explanations that gave less detailed descriptions about the assistant, it was common for responses to say that the assistant ``works out what command you've [said]'' [P169] without reference to how this was done. In other cases, the concept of an algorithm was used as a catch-all term to describe the processing performed by the assistant (``algorithms figure out what is being asked'' [P260]). This suggests that while participants had a functional understanding of what was happening, they lacked the detailed background knowledge required to enrich the explanation.

Others used their own voice assistant as a point of comparison, often subtly widening the scope of their re-explanation to talk about the entire class of devices: ``To activate the voice assistant you must speak clearly, use the voice assistants name to activate and give a command or question'' [P91]. Another strategy employed was to use other systems they were familiar with to encapsulate areas of functionality (``ask it questions like you would to a search engine like Google'' [P25]). Despite this variation over the specifics of how the assistant interpreted requests, nearly every participant identified at least one aspect of the vignettes in their own explanations and very few made incorrect statements about the assistant.

A range of concerns arising from the content of the vignettes was present in participants' re-explanations. People often described how inaccuracies in the voice recognition process might lead to errors when the voice assistant `misheard' an instruction, performing the wrong action or the right action with incorrect parameters. Participants also described the possibility that the assistant would be unable to ``distinguish between what is appropriate for recording and what isn't'' [P360] and start recording when it had not been asked to do anything. Given the similarities in their operation, the effect of reliability explanations on perceptions was often clear: ``I'm not the best at explaining, but [I] would say it's a similar idea to an Alexa but not as trusting. And that I would be worried because of interference that it couldn't work properly.'' [P201]. Despite the fact that all participants owned a voice assistant themselves, several expressed that they would be hesitant to use the assistant described to them, often citing the apparent insecurity or unreliability of the assistant: ``I wouldn't be comfortable using this personally.'' [P81].

In a smaller amount of these cases the errors of the assistant were directly linked to privacy concerns, such as sending personal data to the wrong external party. Those receiving privacy content as part of the offline vignette (0E), often instead mentioned the privacy \textit{benefits} of an assistant that was not connected to the internet. While speaker recognition appeared in some re-explanations from participants who had seen security information, only a small number voiced concerns over the potential for misuse without it. This was the only reason given for concerns over fraud or theft via the assistant---despite being described to a sixth of the participants, no user-given re-explanation mentioned `skill squatting' or skills collecting data without permission.

\section{Discussion: Crafting Explanations for Voice Assistants}\label{sec:discussion}
\subsection{The Presence of First and Third Party Entities}
The results suggest that online functionality is considered to be less trustworthy, private, and secure. On the surface this is unsurprising given that online functionality introduces additional complexity to a system and broadens its attack surface. But the fact that participants distinguished between first and third parties in privacy and security perceptions \textit{was} unexpected given that many concerns identified in the re-explanations focused on local actions (e.g. inappropriate recording and speaker recognition), and the focus on user-device interactions in the security and privacy literature~\cite{10.1145/3412383}. In explaining it we might turn to wider perceptions about connectedness and smartness, where the presence of app stores clearly demarcates the line between the involvement of first and third party entities. Prior work has shown that users identify app stores as potential risks~\cite{10.1145/3357236.3395501}, and it is possible they applied the same logic in the present study; local actions would therefore be concerning because data from them is expected to be shared. This would, however, conflict with other work which found that users often did not mention third parties when describing VA interactions involving third party skills~\cite{abdi2019more}. Given the significant differences found \textit{between} first and third party entities and the fact that this was largely absent from the re-explanations, it seems probable that users do distinguish between first and third party entities even if the associated mental models are not sufficiently detailed to explain why. Overall the nature of online functionality had less of an impact than the way that the assistant was explained, with mean response differences between connectivity conditions almost always lower than between explanation components.

\subsection{The Provision of Trust, Privacy, and Security Information in Explanations}
The results show a stark contrast in the way that the presence of privacy and trust information in explanations affect people's subsequent perceptions of a voice assistant; privacy explanations alleviated privacy concerns, while trust explanations in Q1 aggravated trust concerns. This is also reflected in the large number of trust concerns relative to privacy concerns in re-explanations and the small proportion of concerns over recording which were linked to privacy concerns. In line with previous findings about privacy revelations delivered in isolation~\cite{10.1145/2556288.2557421} (e.g. without an associated increase in understanding or control mechanisms~\cite{10.1145/3313831.3376264}), we had expected that providing participants with privacy explanations would have resulted in them feeling less reassured in the their privacy. It is possible that this was due to the fact that the privacy explanations given were less problematic than participants expected, which would be in line with the misplaced user threat assessments described by the background literature. An alternative explanation is that evidence that privacy risks had been considered was itself enough to assuage concerns. Another important thing to note is that failures of trust in competence are often associated with immediate negative outcomes (such as the assistant not working if disconnected from the internet), whereas failures of privacy are often much less immediately tangible, with longer term effects and the potential for leaked data to be exploited in the future (e.g. information sent to the wrong company then being sold or stolen).

Unlike the response to trust explanations seen in Q1, answers to the other two questions about trust (Questions 2 and 3) did not appear to be significantly impacted by trust explanations at all. Instead, the greatest influence on trust concerns was the brand of assistant that participants owned, followed by the addition of third party functionality. This highlights the unusual relationship of trust when it comes to voice assistants. Conventionally, trust in a manufacturer is related to but distinct from trust in a product---trustworthy manufacturers are more likely to make trustworthy products. However, the complete control that voice assistant manufacturers have over the devices they sell and associated ecosystems means that trust in a voice assistant and trust in its manufacturer should rationally be the same, or a very similar, thing; one cannot trust Alexa without also trusting Amazon, and vice versa. This is also supported by the fact that for the more vendor--focused Q2 and Q3 the significant difference came with the addition of third party functionality, with no such difference between offline and first party functionality.

These differences between users of different devices aligns with findings by \citeauthor{abdi2019more} that voice assistant users are unlikely to have developed mental models beyond their own voice assistant~\cite{abdi2019more}, suggesting that they may be projecting concerns over their own device onto the study vignettes. Interpreting these results also raises a question of causality---do people choose products like Siri over alternatives \textit{because} they had greater general concerns over privacy and security, or do their concerns later align with public perceptions (e.g. valuing privacy more highly because Siri is marketed as being privacy preserving)? That product providers have the greatest effect on trust echoes the results of other recent work on anthropomorphism in voice assistants~\cite{10.1145/3479515}. This raises interesting questions about competition and diversity in the sector, suggesting that new entrants distinguishing themselves with functionality (like the Mycroft assistant being ``private and open''\footnote{\url{https://mycroft.ai/}}) are much less likely to gain traction and become trusted than a voice assistant made by an established brand that people already trust. An exception to this might be moving functionality offline, something that Apple has announced will begin happening with Siri as of iOS 15\footnote{``Siri adds on-device speech recognition, so the audio of your requests is processed on your iPhone or iPad by default. And on-device processing also means Siri can perform many tasks without an internet connection.'' (\url{https://www.apple.com/uk/ios/ios-15-preview/})}.

As described above, there was an underlying incident anxiety factor similar to~\cite{cannizzaro2020trust}, centred around the likelihood of privacy and security breaches, as well as the impact of the latter. Interestingly, when comparing the results to \citeauthor{cannizzaro2020trust} trust in competence (Q1) did not emerge as a significant component in the factor. Further work is needed in order to determine the extent to which this is a result of the shift in focus from smart homes to voice assistants, different participant samples, and/or changes in smart homes between the times that the studies were carried out.

\subsection{Targeting Privacy and Trust Concerns}
At a high level, the participant explanations show a general trend whereby people were able to interpret and understand common behaviours and drawbacks associated with voice assistants, such as the use of wake words and the recording of speech. Problems with devices failing to understand user requests and failing to distinguish requests made by different users were also relatively well described by participants. At the same time, despite the blurred boundaries between participants' own devices and the hypothetical study device, explanation components were almost exclusively given when they were also present in the vignette shown to participants. One interpretation would be that this knowledge is usually backgrounded and/or acquired unconsciously but can be explained on demand: people may not think about these concepts day-to-day, but the vignettes gave them the mental scaffolding to express what they already knew. We hypothesise that where participants were unfamiliar with a concept presented to them in a vignette they were likely to omit it from their subsequent re-explanation, hence the lack of more complex security threats in responses.

At the same time, analysis of participant's perceptions shows that privacy concerns are mitigated by privacy explanations whereas trust/competence explanations exacerbate trust concerns. So if trust explanations require the most improvement from the baseline presented in this study, what is to be done? The incidence rate of key words in the re-explanations suggests that more can be done to improve users' mental models of how voice assistants process speech, but this mainly affects conversational repair. As with privacy concerns where prior work has shown the ineffectiveness of and feelings of dejected acceptance caused by warnings where no remedial action can be taken~\cite{10.1145/2556288.2557421}, designers should be wary of the same trap here; there is little users can do to act on feelings of trust or distrust towards current voice assistants.

As a result, we posit that the competence concerns identified by the study results represent important priorities for voice assistant \textit{manufacturers} to address via software and platform updates. In the interim, we see the use of multimodal interaction as a key tool to mitigate the effects of these concerns. Encouraging the use of design elements such as visual feedback and smartphone confirmation dialogues gives users the ability to verify the correct operation of their assistant in ways that feel familiar and comfortable, while still allowing for convenient hands-free operation most of the time. This approach would also address more widespread problems around low user confidence when shopping via voice assistants~\cite{abdi2019more}. Allowing users to balance connectivity with functionality is another option, and will likely ease concerns even if not ultimately used by the majority of users~\cite{lindley2017internet}.

For privacy concerns, the way forward seems clear. If outlining the extent of privacy risks when using voice assistants reduces the concerns that they have, then providing these explanations is worthwhile. Further work is required to determine exactly why this is the case, as discussed above a possible reason is that people may also be more accustomed to seeing privacy warnings/agreeing to privacy policies which then have little noticeable impact on them, and privacy explanations alone may therefore not be appropriate for increasing knowledge and understanding (see~\cite{10.1145/3313831.3376264}). This points to the adoption of different approaches depending on how well concepts are understood. Quick snippets delivered just-in-time would serve to remind users of important considerations before taking an action (e.g. when linking accounts), whereas more in-depth explanations supported by displays and other devices might be more appropriate for concepts that users are less likely to have an intuitive grasp of (and would therefore be unlikely to acquire from a quick analogy). 

\subsection{Guiding Users with Analogies}
In their explanations, participants readily compared the hypothetical voice assistant in the vignette to their own devices and other digital technologies present in everyday life. A clear next step would be the development of analogies or short explanations that increase understanding about other aspects of how voice assistants work (such as using `building blocks' to represent the parsing of requests), which could be delivered when the device fails to understand the user's intent as a form of conversational grounding~\cite{10.1145/3392838}. This would also be appropriate when targeting poorly understood features or risks, such as the potential for `skill squatting'~\cite{kumar2018skill}. Given the presence of anthropomorphism and potential uses of social reasoning present in the results, there may be a case for social analogies as well as technical ones (e.g. comparing microtransactions within skills/actions to buying a used car in terms of the social mechanics involved).

The prioritisation of explanation concepts is also likely to shift over time. The results of the study suggest that participants, likely primed by experiences with their own voice assistants, saw failings of speech recognition as the greatest concern. But as speech recognition improves and conversation models adapt to enable more natural exchanges (e.g.~\cite{10.1145/3411764.3445640}), helping people better understand the functionality offered by voice assistants will become of greater importance; once the technology becomes sufficiently accurate ``[end-users] do not expect to have that level of insight because they have no practical need of it''~\cite{nilsson2019breaching}.

Indeed, these techniques will become increasingly important as we begin to make assistants that guide people through more complex requests that might include follow-on questions that narrow the scope of an initial enquiry (e.g. styling an assistant as a `coach'~\cite{10.1145/3313831.3376493} or similar). Here these kinds of short, contextual snippets can not only convey the limitations of the assistant, but also the assumptions and boundaries of the dialectic process itself.

\subsection{The Future of Voice Assistant Explanations}
Based on our findings, we ask ourselves the extent to which explanations for voice assistants should focus on reassuring users versus describing the drawbacks of the technology. Going forward we see two main applications for explanations in voice assistants. The first, as demonstrated in the vignettes, is to convey the benefits and drawbacks of contemporary technologies, highlighting relevant information at the point of use that can be further supplemented by other media (e.g. information cards on a display). Following work by \citeauthor{schaub2017designing}~\cite{schaub2017designing}, these explanations should be short and specific, delivered in-context during an interaction. They should also present users with meaningful opportunities to change their actions and/or remedy potential problems. For example, when explaining the potential privacy implications of sending information to a third party skill, offline processing could be offered as an alternative, with the privacy-reliability exchange briefly outlined. We present the further development of these explanations as an open challenge to the research community, and look forward to making our own contributions in future work. 

The second is to convey technical mechanisms put in place to protect users. Given public perceptions and media reports, convincing users as to the efficacy of a protection mechanism is likely to be a challenging task, requiring the use of explanations subject to the caveats given above. For instance, while Amazon has a mechanism to allow users to delete audio recordings associated with previous voice commands, there have been public outcries reported in the media as many people are not aware that this is possible or that recordings are stored\footnote{e.g. ``How to listen to hidden Alexa recordings of your conversations – and then delete them'' (\url{https://www.thesun.co.uk/tech/10987517/listen-alexa-recordings-how-amazon-delete/})}. These types of explanations will become increasingly important as the mechanisms to protect users in voice assistants become more sophisticated. One example would be the introduction of controls over information flows across the voice assistant ecosystem (including to third parties as in our 3E condition), that could come with strong default configurations \cite{abdi2021privacy} and/or learn user preferences over time~\cite{zhan2022model}. For these controls it would be crucial to explain in an accessible way and at the right time what may happen to users' data. A second key example is the introduction of provable privacy and security properties to future voice assistants, constraining the  behaviour of their underlying AI/ML models. Research on the verification and dynamic monitoring of such models is currently a hot topic, with promising results to date~\cite{batten2021efficient, huang2017safety, criado2016selective}.

\subsection{Limitations}\label{sec:limitations}
The greatest limitation of the study lies in the viewpoints and devices of its participants. To reduce complexity respondents were all resident in the UK, and as a result used a small set of devices. In future experiments we aim to further explore the ways in which the perceptions reported here might differ across the globe. Another artefact of the survey design was that some participants were exposed to more concepts than others, and it is possible (though unlikely) that a saturation point existed after which new concepts were not retained. This would manifest in the results as a lower incidence of certain codes amongst groups with longer explanations (e.g. 3E/TPS), but this did not appear to be the case during analysis.

Methodologically, there are also many possible ways to deliver explanations, of which the format presented here is but one. Different approaches to the challenges we discuss will yield different results; the goal of this paper was not to determine the best approach to voice assistant explanations, but rather to determine their general utility and inform their future refinement. By focusing on trust as competence when designing the vignette explanations we were not able to manipulate trust as benevolence or integrity when answering the research questions, and the results suggest that participants subsequently relied on experiences with the vendors of their own devices when answering those questions. The lack of an existing validated questionnaire for the concepts being explored, while mitigated through careful re-use of existing questions, is unfortunate and remains an opportunity for future work.

\section{Conclusion}
Voice assistants occupy an unfortunate meeting point between sleek smart home gadgetry and lower bandwidth voice interfaces. This can make understanding how they work difficult, and makes it harder to assess the benefits and drawbacks of their use. By studying changes in peoples' perceptions of different voice assistants described in various ways, we show the vastly different responses to different kinds of explanations.

Combining this with participants' own re-explanations gave a unique opportunity to see the functionality---and concerns---that were reflected back, highlighting gaps in people's understanding and opening up a discussion on issues that explanations are apt to mitigate, as well as those they are not. The findings suggest that the introduction of online functionality and the absence of privacy explanations both contribute to a general sense of incident anxiety, and that users differentiate and respond to relatively subtle combinations of first and third party entities. The contrasting reactions to privacy and reliability explanations is an important one that can be used to shape the development of future voice assistants and associated explanations, and shows the importance of work by developers towards decreasing misfires and misinterpretations.

Reflecting on the unique role that trust plays in respect to voice assistants, we argue that trust in a voice assistant is logically equivalent to trust in its manufacturer, and lay out the challenges faced by new entrants to the voice assistant market. While the dominance of trust in manufacturers as a predictor of trust in voice assistants may appear to be bad news given the track records of the companies operating in this space, positive responses to offline functionality may yet prove to be good news for everyone (and particularly the privacy concerned).

\begin{acks}
We would like to thank Carlota Vazquez Gonzalez for her assistance with data curation. This work is part of the EPSRC-funded Secure AI Assistants project (grant EP/T026723/1).
\end{acks}

\bibliographystyle{ACM-Reference-Format}
\bibliography{main}


\begin{thebibliography}{96}


\ifx \showCODEN    \undefined \def \showCODEN     #1{\unskip}     \fi
\ifx \showDOI      \undefined \def \showDOI       #1{#1}\fi
\ifx \showISBNx    \undefined \def \showISBNx     #1{\unskip}     \fi
\ifx \showISBNxiii \undefined \def \showISBNxiii  #1{\unskip}     \fi
\ifx \showISSN     \undefined \def \showISSN      #1{\unskip}     \fi
\ifx \showLCCN     \undefined \def \showLCCN      #1{\unskip}     \fi
\ifx \shownote     \undefined \def \shownote      #1{#1}          \fi
\ifx \showarticletitle \undefined \def \showarticletitle #1{#1}   \fi
\ifx \showURL      \undefined \def \showURL       {\relax}        \fi
\providecommand\bibfield[2]{#2}
\providecommand\bibinfo[2]{#2}
\providecommand\natexlab[1]{#1}
\providecommand\showeprint[2][]{arXiv:#2}

\bibitem[\protect\citeauthoryear{Abdi, Ramokapane, and Such}{Abdi
  et~al\mbox{.}}{2019}]%
        {abdi2019more}
\bibfield{author}{\bibinfo{person}{Noura Abdi}, \bibinfo{person}{Kopo
  Ramokapane}, {and} \bibinfo{person}{Jose Such}.}
  \bibinfo{year}{2019}\natexlab{}.
\newblock \showarticletitle{More than smart speakers: security and privacy
  perceptions of smart home personal assistants}. In
  \bibinfo{booktitle}{\emph{Fifteenth Symposium on Usable Privacy and Security
  ($\{$SOUPS$\}$ 2019)}}.
\newblock


\bibitem[\protect\citeauthoryear{Abdi, Zhan, Ramokapane, and Such}{Abdi
  et~al\mbox{.}}{2021}]%
        {abdi2021privacy}
\bibfield{author}{\bibinfo{person}{Noura Abdi}, \bibinfo{person}{Xiao Zhan},
  \bibinfo{person}{Kopo~M Ramokapane}, {and} \bibinfo{person}{Jose Such}.}
  \bibinfo{year}{2021}\natexlab{}.
\newblock \showarticletitle{Privacy Norms for Smart Home Personal Assistants}.
  In \bibinfo{booktitle}{\emph{Proceedings of the 2021 CHI Conference on Human
  Factors in Computing Systems}}. \bibinfo{pages}{1--14}.
\newblock


\bibitem[\protect\citeauthoryear{Abercrombie, Curry, Pandya, and
  Rieser}{Abercrombie et~al\mbox{.}}{2021}]%
        {abercrombie2021alexa}
\bibfield{author}{\bibinfo{person}{Gavin Abercrombie},
  \bibinfo{person}{Amanda~Cercas Curry}, \bibinfo{person}{Mugdha Pandya}, {and}
  \bibinfo{person}{Verena Rieser}.} \bibinfo{year}{2021}\natexlab{}.
\newblock \bibinfo{title}{Alexa, Google, Siri: What are Your Pronouns? Gender
  and Anthropomorphism in the Design and Perception of Conversational
  Assistants}.
\newblock
\newblock
\showeprint[arxiv]{2106.02578}~[cs.AI]


\bibitem[\protect\citeauthoryear{Acar, Fereidooni, Abera, Sikder, Miettinen,
  Aksu, Conti, Sadeghi, and Uluagac}{Acar et~al\mbox{.}}{2020}]%
        {10.1145/3395351.3399421}
\bibfield{author}{\bibinfo{person}{Abbas Acar}, \bibinfo{person}{Hossein
  Fereidooni}, \bibinfo{person}{Tigist Abera}, \bibinfo{person}{Amit~Kumar
  Sikder}, \bibinfo{person}{Markus Miettinen}, \bibinfo{person}{Hidayet Aksu},
  \bibinfo{person}{Mauro Conti}, \bibinfo{person}{Ahmad-Reza Sadeghi}, {and}
  \bibinfo{person}{Selcuk Uluagac}.} \bibinfo{year}{2020}\natexlab{}.
\newblock \showarticletitle{Peek-a-Boo: I See Your Smart Home Activities, Even
  Encrypted!}. In \bibinfo{booktitle}{\emph{Proceedings of the 13th ACM
  Conference on Security and Privacy in Wireless and Mobile Networks}} (Linz,
  Austria) \emph{(\bibinfo{series}{WiSec '20})}.
  \bibinfo{publisher}{Association for Computing Machinery},
  \bibinfo{address}{New York, NY, USA}, \bibinfo{pages}{207–218}.
\newblock
\showISBNx{9781450380065}
\urldef\tempurl%
\url{https://doi.org/10.1145/3395351.3399421}
\showDOI{\tempurl}


\bibitem[\protect\citeauthoryear{Andow, Mahmud, Wang, Whitaker, Enck, Reaves,
  Singh, and Xie}{Andow et~al\mbox{.}}{2019}]%
        {andow2019policylint}
\bibfield{author}{\bibinfo{person}{Benjamin Andow},
  \bibinfo{person}{Samin~Yaseer Mahmud}, \bibinfo{person}{Wenyu Wang},
  \bibinfo{person}{Justin Whitaker}, \bibinfo{person}{William Enck},
  \bibinfo{person}{Bradley Reaves}, \bibinfo{person}{Kapil Singh}, {and}
  \bibinfo{person}{Tao Xie}.} \bibinfo{year}{2019}\natexlab{}.
\newblock \showarticletitle{Policylint: investigating internal privacy policy
  contradictions on Google play}. In \bibinfo{booktitle}{\emph{28th
  $\{$USENIX$\}$ Security Symposium ($\{$USENIX$\}$ Security 19)}}.
  \bibinfo{pages}{585--602}.
\newblock


\bibitem[\protect\citeauthoryear{Axtell and Munteanu}{Axtell and
  Munteanu}{2021}]%
        {10.1145/3411764.3445640}
\bibfield{author}{\bibinfo{person}{Benett Axtell} {and} \bibinfo{person}{Cosmin
  Munteanu}.} \bibinfo{year}{2021}\natexlab{}.
\newblock \showarticletitle{Tea, Earl Grey, Hot: Designing Speech Interactions
  from the Imagined Ideal of Star Trek}. In
  \bibinfo{booktitle}{\emph{Proceedings of the 2021 CHI Conference on Human
  Factors in Computing Systems}} (Yokohama, Japan) \emph{(\bibinfo{series}{CHI
  '21})}. \bibinfo{publisher}{Association for Computing Machinery},
  \bibinfo{address}{New York, NY, USA}, Article \bibinfo{articleno}{249},
  \bibinfo{numpages}{14}~pages.
\newblock
\showISBNx{9781450380966}
\urldef\tempurl%
\url{https://doi.org/10.1145/3411764.3445640}
\showDOI{\tempurl}


\bibitem[\protect\citeauthoryear{Batten, Kouvaros, Lomuscio, and Zheng}{Batten
  et~al\mbox{.}}{2021}]%
        {batten2021efficient}
\bibfield{author}{\bibinfo{person}{Ben Batten}, \bibinfo{person}{Panagiotis
  Kouvaros}, \bibinfo{person}{Alessio Lomuscio}, {and} \bibinfo{person}{Yang
  Zheng}.} \bibinfo{year}{2021}\natexlab{}.
\newblock \showarticletitle{Efficient neural network verification via
  layer-based semidefinite relaxations and linear cuts}. In
  \bibinfo{booktitle}{\emph{30th International Joint Conference on Artificial
  Intelligence (IJCAI-21), accepted}}.
\newblock


\bibitem[\protect\citeauthoryear{Braun and Clarke}{Braun and Clarke}{2012}]%
        {braun2012thematic}
\bibfield{author}{\bibinfo{person}{Virginia Braun} {and}
  \bibinfo{person}{Victoria Clarke}.} \bibinfo{year}{2012}\natexlab{}.
\newblock \showarticletitle{Thematic analysis.}
\newblock  (\bibinfo{year}{2012}).
\newblock


\bibitem[\protect\citeauthoryear{Burgess}{Burgess}{2017}]%
        {burgess2017google}
\bibfield{author}{\bibinfo{person}{Matt Burgess}.}
  \bibinfo{year}{2017}\natexlab{}.
\newblock \bibinfo{booktitle}{\emph{Google stops 'What is the Whopper burger?'
  ad triggering Google Home}}.
\newblock
\urldef\tempurl%
\url{https://www.wired.co.uk/article/google-home-burger-king-ad}
\showURL{%
\tempurl}


\bibitem[\protect\citeauthoryear{Cannizzaro, Procter, Ma, and Maple}{Cannizzaro
  et~al\mbox{.}}{2020}]%
        {cannizzaro2020trust}
\bibfield{author}{\bibinfo{person}{Sara Cannizzaro}, \bibinfo{person}{Rob
  Procter}, \bibinfo{person}{Sinong Ma}, {and} \bibinfo{person}{Carsten
  Maple}.} \bibinfo{year}{2020}\natexlab{}.
\newblock \showarticletitle{Trust in the smart home: Findings from a nationally
  representative survey in the UK}.
\newblock \bibinfo{journal}{\emph{PLOS ONE}} \bibinfo{volume}{15},
  \bibinfo{number}{5} (\bibinfo{date}{05} \bibinfo{year}{2020}),
  \bibinfo{pages}{1--30}.
\newblock
\urldef\tempurl%
\url{https://doi.org/10.1371/journal.pone.0231615}
\showDOI{\tempurl}


\bibitem[\protect\citeauthoryear{Chiang, Chang, Chuang, Chou, Lee, Lin,
  Jiang~Chen, and Chang}{Chiang et~al\mbox{.}}{2020}]%
        {chiang2020exploring}
\bibfield{author}{\bibinfo{person}{Yi-Shyuan Chiang}, \bibinfo{person}{Ruei-Che
  Chang}, \bibinfo{person}{Yi-Lin Chuang}, \bibinfo{person}{Shih-Ya Chou},
  \bibinfo{person}{Hao-Ping Lee}, \bibinfo{person}{I-Ju Lin},
  \bibinfo{person}{Jian-Hua Jiang~Chen}, {and} \bibinfo{person}{Yung-Ju
  Chang}.} \bibinfo{year}{2020}\natexlab{}.
\newblock \showarticletitle{Exploring the design space of user-system
  communication for smart-home routine assistants}. In
  \bibinfo{booktitle}{\emph{Proceedings of the 2020 CHI Conference on Human
  Factors in Computing Systems}}. \bibinfo{pages}{1--14}.
\newblock


\bibitem[\protect\citeauthoryear{Cho, Sundar, Abdullah, and Motalebi}{Cho
  et~al\mbox{.}}{2020}]%
        {10.1145/3313831.3376551}
\bibfield{author}{\bibinfo{person}{Eugene Cho}, \bibinfo{person}{S.~Shyam
  Sundar}, \bibinfo{person}{Saeed Abdullah}, {and} \bibinfo{person}{Nasim
  Motalebi}.} \bibinfo{year}{2020}\natexlab{}.
\newblock \bibinfo{booktitle}{\emph{Will Deleting History Make Alexa More
  Trustworthy? Effects of Privacy and Content Customization on User Experience
  of Smart Speakers}}.
\newblock \bibinfo{publisher}{Association for Computing Machinery},
  \bibinfo{address}{New York, NY, USA}, \bibinfo{pages}{1–13}.
\newblock
\showISBNx{9781450367080}
\urldef\tempurl%
\url{https://doi.org/10.1145/3313831.3376551}
\showURL{%
\tempurl}


\bibitem[\protect\citeauthoryear{Cho and Rader}{Cho and Rader}{2020}]%
        {10.1145/3392838}
\bibfield{author}{\bibinfo{person}{Janghee Cho} {and} \bibinfo{person}{Emilee
  Rader}.} \bibinfo{year}{2020}\natexlab{}.
\newblock \showarticletitle{The Role of Conversational Grounding in Supporting
  Symbiosis Between People and Digital Assistants}.
\newblock \bibinfo{journal}{\emph{Proc. ACM Hum.-Comput. Interact.}}
  \bibinfo{volume}{4}, \bibinfo{number}{CSCW1}, Article
  \bibinfo{articleno}{033} (\bibinfo{date}{May} \bibinfo{year}{2020}),
  \bibinfo{numpages}{28}~pages.
\newblock
\urldef\tempurl%
\url{https://doi.org/10.1145/3392838}
\showDOI{\tempurl}


\bibitem[\protect\citeauthoryear{Clark, Newman, and Dutta}{Clark
  et~al\mbox{.}}{2017}]%
        {clark2017devices}
\bibfield{author}{\bibinfo{person}{Meghan Clark}, \bibinfo{person}{Mark~W.
  Newman}, {and} \bibinfo{person}{Prabal Dutta}.}
  \bibinfo{year}{2017}\natexlab{}.
\newblock \showarticletitle{Devices and Data and Agents, Oh My: How Smart Home
  Abstractions Prime End-User Mental Models}.
\newblock \bibinfo{journal}{\emph{Proc. ACM Interact. Mob. Wearable Ubiquitous
  Technol.}} \bibinfo{volume}{1}, \bibinfo{number}{3}, Article
  \bibinfo{articleno}{44} (\bibinfo{date}{Sept.} \bibinfo{year}{2017}),
  \bibinfo{numpages}{26}~pages.
\newblock
\urldef\tempurl%
\url{https://doi.org/10.1145/3132031}
\showDOI{\tempurl}


\bibitem[\protect\citeauthoryear{Cohn-Gordon, Cremers, Dowling, Garratt, and
  Stebila}{Cohn-Gordon et~al\mbox{.}}{2020}]%
        {cohn2020formal}
\bibfield{author}{\bibinfo{person}{Katriel Cohn-Gordon}, \bibinfo{person}{Cas
  Cremers}, \bibinfo{person}{Benjamin Dowling}, \bibinfo{person}{Luke Garratt},
  {and} \bibinfo{person}{Douglas Stebila}.} \bibinfo{year}{2020}\natexlab{}.
\newblock \showarticletitle{A formal security analysis of the signal messaging
  protocol}.
\newblock \bibinfo{journal}{\emph{Journal of Cryptology}} \bibinfo{volume}{33},
  \bibinfo{number}{4} (\bibinfo{year}{2020}), \bibinfo{pages}{1914--1983}.
\newblock


\bibitem[\protect\citeauthoryear{Coulton and Lindley}{Coulton and
  Lindley}{2019}]%
        {coulton2019more}
\bibfield{author}{\bibinfo{person}{Paul Coulton} {and}
  \bibinfo{person}{Joseph~Galen Lindley}.} \bibinfo{year}{2019}\natexlab{}.
\newblock \showarticletitle{More-than human centred design: Considering other
  things}.
\newblock \bibinfo{journal}{\emph{The Design Journal}} \bibinfo{volume}{22},
  \bibinfo{number}{4} (\bibinfo{year}{2019}), \bibinfo{pages}{463--481}.
\newblock


\bibitem[\protect\citeauthoryear{Cowan, Branigan, Begum, McKenna, and
  Szekely}{Cowan et~al\mbox{.}}{2017a}]%
        {cowan2017they}
\bibfield{author}{\bibinfo{person}{Benjamin~R Cowan}, \bibinfo{person}{Holly~P
  Branigan}, \bibinfo{person}{Habiba Begum}, \bibinfo{person}{Lucy McKenna},
  {and} \bibinfo{person}{Eva Szekely}.} \bibinfo{year}{2017}\natexlab{a}.
\newblock \showarticletitle{They Know as Much as We Do: Knowledge Estimation
  and Partner Modelling of Artificial Partners.}. In
  \bibinfo{booktitle}{\emph{CogSci}}.
\newblock


\bibitem[\protect\citeauthoryear{Cowan, Pantidi, Coyle, Morrissey, Clarke,
  Al-Shehri, Earley, and Bandeira}{Cowan et~al\mbox{.}}{2017b}]%
        {10.1145/3098279.3098539}
\bibfield{author}{\bibinfo{person}{Benjamin~R. Cowan}, \bibinfo{person}{Nadia
  Pantidi}, \bibinfo{person}{David Coyle}, \bibinfo{person}{Kellie Morrissey},
  \bibinfo{person}{Peter Clarke}, \bibinfo{person}{Sara Al-Shehri},
  \bibinfo{person}{David Earley}, {and} \bibinfo{person}{Natasha Bandeira}.}
  \bibinfo{year}{2017}\natexlab{b}.
\newblock \showarticletitle{"What Can i Help You with?": Infrequent Users'
  Experiences of Intelligent Personal Assistants}. In
  \bibinfo{booktitle}{\emph{Proceedings of the 19th International Conference on
  Human-Computer Interaction with Mobile Devices and Services}} (Vienna,
  Austria) \emph{(\bibinfo{series}{MobileHCI '17})}.
  \bibinfo{publisher}{Association for Computing Machinery},
  \bibinfo{address}{New York, NY, USA}, Article \bibinfo{articleno}{43},
  \bibinfo{numpages}{12}~pages.
\newblock
\showISBNx{9781450350754}
\urldef\tempurl%
\url{https://doi.org/10.1145/3098279.3098539}
\showDOI{\tempurl}


\bibitem[\protect\citeauthoryear{Criado and Such}{Criado and Such}{2016}]%
        {criado2016selective}
\bibfield{author}{\bibinfo{person}{Natalia Criado} {and} \bibinfo{person}{Jose
  Such}.} \bibinfo{year}{2016}\natexlab{}.
\newblock \showarticletitle{Selective norm monitoring}. In
  \bibinfo{booktitle}{\emph{Proceedings of the Twenty-Fifth International Joint
  Conference on Artificial Intelligence (IJCAI-16)}}. AAAI Press,
  \bibinfo{pages}{208--214}.
\newblock


\bibitem[\protect\citeauthoryear{Dambanemuya and Diakopoulos}{Dambanemuya and
  Diakopoulos}{2020}]%
        {dambanemuya2020alexa}
\bibfield{author}{\bibinfo{person}{Henry~K Dambanemuya} {and}
  \bibinfo{person}{Nicholas Diakopoulos}.} \bibinfo{year}{2020}\natexlab{}.
\newblock \showarticletitle{“Alexa, what is going on with the impeachment?”
  Evaluating smart speakers for news quality}. In
  \bibinfo{booktitle}{\emph{Proc. Computation+ Journalism Symposium}}.
\newblock


\bibitem[\protect\citeauthoryear{Danezis, Domingo-Ferrer, Hansen, Hoepman,
  Metayer, Tirtea, and Schiffner}{Danezis et~al\mbox{.}}{2014}]%
        {danezis2015privacy}
\bibfield{author}{\bibinfo{person}{G. Danezis}, \bibinfo{person}{J.
  Domingo-Ferrer}, \bibinfo{person}{M. Hansen}, \bibinfo{person}{J. Hoepman},
  \bibinfo{person}{D. Metayer}, \bibinfo{person}{R. Tirtea}, {and}
  \bibinfo{person}{S. Schiffner}.} \bibinfo{year}{2014}\natexlab{}.
\newblock \showarticletitle{Privacy and Data Protection by Design-from policy
  to engineering}.
\newblock \bibinfo{journal}{\emph{ENISA}} (\bibinfo{year}{2014}).
\newblock


\bibitem[\protect\citeauthoryear{DeVito, Gergle, and Birnholtz}{DeVito
  et~al\mbox{.}}{2017}]%
        {10.1145/3025453.3025659}
\bibfield{author}{\bibinfo{person}{Michael~A. DeVito}, \bibinfo{person}{Darren
  Gergle}, {and} \bibinfo{person}{Jeremy Birnholtz}.}
  \bibinfo{year}{2017}\natexlab{}.
\newblock \bibinfo{booktitle}{\emph{"Algorithms Ruin Everything": \#RIPTwitter,
  Folk Theories, and Resistance to Algorithmic Change in Social Media}}.
\newblock \bibinfo{publisher}{Association for Computing Machinery},
  \bibinfo{address}{New York, NY, USA}, \bibinfo{pages}{3163–3174}.
\newblock
\showISBNx{9781450346559}
\urldef\tempurl%
\url{https://doi.org/10.1145/3025453.3025659}
\showURL{%
\tempurl}


\bibitem[\protect\citeauthoryear{Doyle, Clark, and Cowan}{Doyle
  et~al\mbox{.}}{2021}]%
        {10.1145/3411764.3445206}
\bibfield{author}{\bibinfo{person}{Philip~R Doyle}, \bibinfo{person}{Leigh
  Clark}, {and} \bibinfo{person}{Benjamin~R. Cowan}.}
  \bibinfo{year}{2021}\natexlab{}.
\newblock \bibinfo{booktitle}{\emph{What Do We See in Them? Identifying
  Dimensions of Partner Models for Speech Interfaces Using a Psycholexical
  Approach}}.
\newblock \bibinfo{publisher}{Association for Computing Machinery},
  \bibinfo{address}{New York, NY, USA}.
\newblock
\showISBNx{9781450380966}
\urldef\tempurl%
\url{https://doi.org/10.1145/3411764.3445206}
\showURL{%
\tempurl}


\bibitem[\protect\citeauthoryear{Dunin-Underwood}{Dunin-Underwood}{2020}]%
        {dunin2020alexa}
\bibfield{author}{\bibinfo{person}{Anna Dunin-Underwood}.}
  \bibinfo{year}{2020}\natexlab{}.
\newblock \showarticletitle{Alexa, can you keep a secret? Applicability of the
  third-party doctrine to information collected in the home by virtual
  assistants}.
\newblock \bibinfo{journal}{\emph{Information \& Communications Technology
  Law}} \bibinfo{volume}{29}, \bibinfo{number}{1} (\bibinfo{year}{2020}),
  \bibinfo{pages}{101--119}.
\newblock
\urldef\tempurl%
\url{https://doi.org/10.1080/13600834.2020.1676956}
\showDOI{\tempurl}


\bibitem[\protect\citeauthoryear{Edu, Ferrer-Aran, Such, and Suarez-Tangi}{Edu
  et~al\mbox{.}}{2021}]%
        {edu2021skillvet}
\bibfield{author}{\bibinfo{person}{Jide Edu}, \bibinfo{person}{Xavier
  Ferrer-Aran}, \bibinfo{person}{Jose Such}, {and} \bibinfo{person}{Guillermo
  Suarez-Tangi}.} \bibinfo{year}{2021}\natexlab{}.
\newblock \showarticletitle{SkillVet: Automated Traceability Analysis of Amazon
  Alexa Skills}.
\newblock \bibinfo{journal}{\emph{IEEE Transactions on Dependable and Secure
  Computing (TDSC)}} (\bibinfo{year}{2021}).
\newblock


\bibitem[\protect\citeauthoryear{Edu, {Ferrer Aran}, Such, and
  Suarez-Tangil}{Edu et~al\mbox{.}}{2022}]%
        {edu2022measuring}
\bibfield{author}{\bibinfo{person}{Jide Edu}, \bibinfo{person}{Xavier {Ferrer
  Aran}}, \bibinfo{person}{Jose Such}, {and} \bibinfo{person}{Guillermo
  Suarez-Tangil}.} \bibinfo{year}{2022}\natexlab{}.
\newblock \bibinfo{booktitle}{\emph{Measuring Alexa Skill Privacy Practices
  across Three Years}}.
\newblock \bibinfo{publisher}{ACM}, \bibinfo{pages}{670--680}.
\newblock


\bibitem[\protect\citeauthoryear{Edu, Such, and Suarez-Tangil}{Edu
  et~al\mbox{.}}{2020}]%
        {10.1145/3412383}
\bibfield{author}{\bibinfo{person}{Jide Edu}, \bibinfo{person}{Jose Such},
  {and} \bibinfo{person}{Guillermo Suarez-Tangil}.}
  \bibinfo{year}{2020}\natexlab{}.
\newblock \showarticletitle{Smart Home Personal Assistants: A Security and
  Privacy Review}.
\newblock \bibinfo{journal}{\emph{ACM Comput. Surv.}} \bibinfo{volume}{53},
  \bibinfo{number}{6}, Article \bibinfo{articleno}{116} (\bibinfo{date}{Dec.}
  \bibinfo{year}{2020}), \bibinfo{numpages}{36}~pages.
\newblock
\showISSN{0360-0300}
\urldef\tempurl%
\url{https://doi.org/10.1145/3412383}
\showDOI{\tempurl}


\bibitem[\protect\citeauthoryear{Emami-Naeini, Dixon, Agarwal, and
  Cranor}{Emami-Naeini et~al\mbox{.}}{2019}]%
        {10.1145/3290605.3300764}
\bibfield{author}{\bibinfo{person}{Pardis Emami-Naeini}, \bibinfo{person}{Henry
  Dixon}, \bibinfo{person}{Yuvraj Agarwal}, {and} \bibinfo{person}{Lorrie~Faith
  Cranor}.} \bibinfo{year}{2019}\natexlab{}.
\newblock \bibinfo{booktitle}{\emph{Exploring How Privacy and Security Factor
  into IoT Device Purchase Behavior}}.
\newblock \bibinfo{publisher}{Association for Computing Machinery},
  \bibinfo{address}{New York, NY, USA}, \bibinfo{pages}{1–12}.
\newblock
\showISBNx{9781450359702}
\urldef\tempurl%
\url{https://doi.org/10.1145/3290605.3300764}
\showURL{%
\tempurl}


\bibitem[\protect\citeauthoryear{Eslami, Karahalios, Sandvig, Vaccaro, Rickman,
  Hamilton, and Kirlik}{Eslami et~al\mbox{.}}{2016}]%
        {eslami2016first}
\bibfield{author}{\bibinfo{person}{Motahhare Eslami}, \bibinfo{person}{Karrie
  Karahalios}, \bibinfo{person}{Christian Sandvig}, \bibinfo{person}{Kristen
  Vaccaro}, \bibinfo{person}{Aimee Rickman}, \bibinfo{person}{Kevin Hamilton},
  {and} \bibinfo{person}{Alex Kirlik}.} \bibinfo{year}{2016}\natexlab{}.
\newblock \showarticletitle{First I" like" it, then I hide it: Folk Theories of
  Social Feeds}. In \bibinfo{booktitle}{\emph{Proceedings of the 2016 cHI
  conference on human factors in computing systems}}.
  \bibinfo{pages}{2371--2382}.
\newblock


\bibitem[\protect\citeauthoryear{Faklaris, Dabbish, and Hong}{Faklaris
  et~al\mbox{.}}{2019}]%
        {faklaris2019self}
\bibfield{author}{\bibinfo{person}{Cori Faklaris}, \bibinfo{person}{Laura~A
  Dabbish}, {and} \bibinfo{person}{Jason~I Hong}.}
  \bibinfo{year}{2019}\natexlab{}.
\newblock \showarticletitle{A self-report measure of end-user security
  attitudes (SA-6)}. In \bibinfo{booktitle}{\emph{Fifteenth Symposium on Usable
  Privacy and Security ($\{$SOUPS$\}$ 2019)}}.
\newblock


\bibitem[\protect\citeauthoryear{Fischer, Reeves, Porcheron, and
  Sikveland}{Fischer et~al\mbox{.}}{2019}]%
        {10.1145/3342775.3342788}
\bibfield{author}{\bibinfo{person}{Joel~E. Fischer}, \bibinfo{person}{Stuart
  Reeves}, \bibinfo{person}{Martin Porcheron}, {and} \bibinfo{person}{Rein~Ove
  Sikveland}.} \bibinfo{year}{2019}\natexlab{}.
\newblock \showarticletitle{Progressivity for Voice Interface Design}. In
  \bibinfo{booktitle}{\emph{Proceedings of the 1st International Conference on
  Conversational User Interfaces}} (Dublin, Ireland)
  \emph{(\bibinfo{series}{CUI '19})}. \bibinfo{publisher}{Association for
  Computing Machinery}, \bibinfo{address}{New York, NY, USA}, Article
  \bibinfo{articleno}{26}, \bibinfo{numpages}{8}~pages.
\newblock
\showISBNx{9781450371872}
\urldef\tempurl%
\url{https://doi.org/10.1145/3342775.3342788}
\showDOI{\tempurl}


\bibitem[\protect\citeauthoryear{Fleiss, Levin, and Paik}{Fleiss
  et~al\mbox{.}}{2013}]%
        {fleiss2013statistical}
\bibfield{author}{\bibinfo{person}{Joseph~L Fleiss}, \bibinfo{person}{Bruce
  Levin}, {and} \bibinfo{person}{Myunghee~Cho Paik}.}
  \bibinfo{year}{2013}\natexlab{}.
\newblock \bibinfo{booktitle}{\emph{Statistical methods for rates and
  proportions}}.
\newblock \bibinfo{publisher}{john wiley \& sons}.
\newblock


\bibitem[\protect\citeauthoryear{Fruchter and Liccardi}{Fruchter and
  Liccardi}{2018}]%
        {10.1145/3170427.3188448}
\bibfield{author}{\bibinfo{person}{Nathaniel Fruchter} {and}
  \bibinfo{person}{Ilaria Liccardi}.} \bibinfo{year}{2018}\natexlab{}.
\newblock \showarticletitle{Consumer Attitudes Towards Privacy and Security in
  Home Assistants}. In \bibinfo{booktitle}{\emph{Extended Abstracts of the 2018
  CHI Conference on Human Factors in Computing Systems}} (Montreal QC, Canada)
  \emph{(\bibinfo{series}{CHI EA '18})}. \bibinfo{publisher}{Association for
  Computing Machinery}, \bibinfo{address}{New York, NY, USA},
  \bibinfo{pages}{1–6}.
\newblock
\showISBNx{9781450356213}
\urldef\tempurl%
\url{https://doi.org/10.1145/3170427.3188448}
\showDOI{\tempurl}


\bibitem[\protect\citeauthoryear{G{\"u}rses and Del~Alamo}{G{\"u}rses and
  Del~Alamo}{2016}]%
        {gurses2016privacy}
\bibfield{author}{\bibinfo{person}{Seda G{\"u}rses} {and}
  \bibinfo{person}{Jose~M Del~Alamo}.} \bibinfo{year}{2016}\natexlab{}.
\newblock \showarticletitle{Privacy engineering: Shaping an emerging field of
  research and practice}.
\newblock \bibinfo{journal}{\emph{IEEE Security \& Privacy}}
  \bibinfo{volume}{14}, \bibinfo{number}{2} (\bibinfo{year}{2016}),
  \bibinfo{pages}{40--46}.
\newblock


\bibitem[\protect\citeauthoryear{Hoffman, Mueller, Klein, and Litman}{Hoffman
  et~al\mbox{.}}{2018}]%
        {hoffman2018metrics}
\bibfield{author}{\bibinfo{person}{Robert~R Hoffman}, \bibinfo{person}{Shane~T
  Mueller}, \bibinfo{person}{Gary Klein}, {and} \bibinfo{person}{Jordan
  Litman}.} \bibinfo{year}{2018}\natexlab{}.
\newblock \showarticletitle{Metrics for explainable AI: Challenges and
  prospects}.
\newblock \bibinfo{journal}{\emph{arXiv preprint arXiv:1812.04608}}
  (\bibinfo{year}{2018}).
\newblock


\bibitem[\protect\citeauthoryear{Huang, Apthorpe, Li, Acar, and Feamster}{Huang
  et~al\mbox{.}}{2020a}]%
        {10.1145/3397333}
\bibfield{author}{\bibinfo{person}{Danny~Yuxing Huang}, \bibinfo{person}{Noah
  Apthorpe}, \bibinfo{person}{Frank Li}, \bibinfo{person}{Gunes Acar}, {and}
  \bibinfo{person}{Nick Feamster}.} \bibinfo{year}{2020}\natexlab{a}.
\newblock \showarticletitle{IoT Inspector: Crowdsourcing Labeled Network
  Traffic from Smart Home Devices at Scale}.
\newblock \bibinfo{journal}{\emph{Proc. ACM Interact. Mob. Wearable Ubiquitous
  Technol.}} \bibinfo{volume}{4}, \bibinfo{number}{2}, Article
  \bibinfo{articleno}{46} (\bibinfo{date}{June} \bibinfo{year}{2020}),
  \bibinfo{numpages}{21}~pages.
\newblock
\urldef\tempurl%
\url{https://doi.org/10.1145/3397333}
\showDOI{\tempurl}


\bibitem[\protect\citeauthoryear{Huang, Kwiatkowska, Wang, and Wu}{Huang
  et~al\mbox{.}}{2017}]%
        {huang2017safety}
\bibfield{author}{\bibinfo{person}{Xiaowei Huang}, \bibinfo{person}{Marta
  Kwiatkowska}, \bibinfo{person}{Sen Wang}, {and} \bibinfo{person}{Min Wu}.}
  \bibinfo{year}{2017}\natexlab{}.
\newblock \showarticletitle{Safety verification of deep neural networks}. In
  \bibinfo{booktitle}{\emph{International conference on computer aided
  verification}}. Springer, \bibinfo{pages}{3--29}.
\newblock


\bibitem[\protect\citeauthoryear{Huang, Obada-Obieh, and Beznosov}{Huang
  et~al\mbox{.}}{2020b}]%
        {10.1145/3313831.3376529}
\bibfield{author}{\bibinfo{person}{Yue Huang}, \bibinfo{person}{Borke
  Obada-Obieh}, {and} \bibinfo{person}{Konstantin~(Kosta) Beznosov}.}
  \bibinfo{year}{2020}\natexlab{b}.
\newblock \bibinfo{booktitle}{\emph{Amazon vs. My Brother: How Users of Shared
  Smart Speakers Perceive and Cope with Privacy Risks}}.
\newblock \bibinfo{publisher}{Association for Computing Machinery},
  \bibinfo{address}{New York, NY, USA}, \bibinfo{pages}{1–13}.
\newblock
\showISBNx{9781450367080}
\urldef\tempurl%
\url{https://doi.org/10.1145/3313831.3376529}
\showURL{%
\tempurl}


\bibitem[\protect\citeauthoryear{Jensen and Potts}{Jensen and Potts}{2004}]%
        {10.1145/985692.985752}
\bibfield{author}{\bibinfo{person}{Carlos Jensen} {and} \bibinfo{person}{Colin
  Potts}.} \bibinfo{year}{2004}\natexlab{}.
\newblock \bibinfo{booktitle}{\emph{Privacy Policies as Decision-Making Tools:
  An Evaluation of Online Privacy Notices}}.
\newblock \bibinfo{publisher}{Association for Computing Machinery},
  \bibinfo{address}{New York, NY, USA}, \bibinfo{pages}{471–478}.
\newblock
\showISBNx{1581137028}
\urldef\tempurl%
\url{https://doi.org/10.1145/985692.985752}
\showURL{%
\tempurl}


\bibitem[\protect\citeauthoryear{Kaiser and Rice}{Kaiser and Rice}{1974}]%
        {kaiser1974little}
\bibfield{author}{\bibinfo{person}{Henry~F Kaiser} {and} \bibinfo{person}{John
  Rice}.} \bibinfo{year}{1974}\natexlab{}.
\newblock \showarticletitle{Little jiffy, mark IV}.
\newblock \bibinfo{journal}{\emph{Educational and psychological measurement}}
  \bibinfo{volume}{34}, \bibinfo{number}{1} (\bibinfo{year}{1974}),
  \bibinfo{pages}{111--117}.
\newblock


\bibitem[\protect\citeauthoryear{Kang, Dabbish, Fruchter, and Kiesler}{Kang
  et~al\mbox{.}}{2015}]%
        {kang2015my}
\bibfield{author}{\bibinfo{person}{Ruogu Kang}, \bibinfo{person}{Laura
  Dabbish}, \bibinfo{person}{Nathaniel Fruchter}, {and} \bibinfo{person}{Sara
  Kiesler}.} \bibinfo{year}{2015}\natexlab{}.
\newblock \showarticletitle{“my data just goes everywhere:” user mental
  models of the internet and implications for privacy and security}. In
  \bibinfo{booktitle}{\emph{Eleventh Symposium On Usable Privacy and Security
  ($\{$SOUPS$\}$ 2015)}}. \bibinfo{pages}{39--52}.
\newblock


\bibitem[\protect\citeauthoryear{Kannampallil, Smyth, Jones, Payne, and
  Ma}{Kannampallil et~al\mbox{.}}{2020}]%
        {kannampallil2020cognitive}
\bibfield{author}{\bibinfo{person}{Thomas Kannampallil},
  \bibinfo{person}{Joshua~M Smyth}, \bibinfo{person}{Steve Jones},
  \bibinfo{person}{Philip~RO Payne}, {and} \bibinfo{person}{Jun Ma}.}
  \bibinfo{year}{2020}\natexlab{}.
\newblock \showarticletitle{Cognitive plausibility in voice-based AI health
  counselors}.
\newblock \bibinfo{journal}{\emph{NPJ digital medicine}} \bibinfo{volume}{3},
  \bibinfo{number}{1} (\bibinfo{year}{2020}), \bibinfo{pages}{1--4}.
\newblock


\bibitem[\protect\citeauthoryear{Kelley, Bresee, Cranor, and Reeder}{Kelley
  et~al\mbox{.}}{2009}]%
        {10.1145/1572532.1572538}
\bibfield{author}{\bibinfo{person}{Patrick~Gage Kelley},
  \bibinfo{person}{Joanna Bresee}, \bibinfo{person}{Lorrie~Faith Cranor}, {and}
  \bibinfo{person}{Robert~W. Reeder}.} \bibinfo{year}{2009}\natexlab{}.
\newblock \showarticletitle{A "Nutrition Label" for Privacy}. In
  \bibinfo{booktitle}{\emph{Proceedings of the 5th Symposium on Usable Privacy
  and Security}} (Mountain View, California, USA) \emph{(\bibinfo{series}{SOUPS
  '09})}. \bibinfo{publisher}{Association for Computing Machinery},
  \bibinfo{address}{New York, NY, USA}, Article \bibinfo{articleno}{4},
  \bibinfo{numpages}{12}~pages.
\newblock
\showISBNx{9781605587363}
\urldef\tempurl%
\url{https://doi.org/10.1145/1572532.1572538}
\showDOI{\tempurl}


\bibitem[\protect\citeauthoryear{Klein, Elphinstone, Heiser, Andronick, Cock,
  Derrin, Elkaduwe, Engelhardt, Kolanski, Norrish, et~al\mbox{.}}{Klein
  et~al\mbox{.}}{2009}]%
        {klein2009sel4}
\bibfield{author}{\bibinfo{person}{Gerwin Klein}, \bibinfo{person}{Kevin
  Elphinstone}, \bibinfo{person}{Gernot Heiser}, \bibinfo{person}{June
  Andronick}, \bibinfo{person}{David Cock}, \bibinfo{person}{Philip Derrin},
  \bibinfo{person}{Dhammika Elkaduwe}, \bibinfo{person}{Kai Engelhardt},
  \bibinfo{person}{Rafal Kolanski}, \bibinfo{person}{Michael Norrish},
  {et~al\mbox{.}}} \bibinfo{year}{2009}\natexlab{}.
\newblock \showarticletitle{seL4: Formal verification of an OS kernel}. In
  \bibinfo{booktitle}{\emph{Proceedings of the ACM SIGOPS 22nd symposium on
  Operating systems principles}}. \bibinfo{pages}{207--220}.
\newblock


\bibitem[\protect\citeauthoryear{Koo, Kwac, Ju, Steinert, Leifer, and Nass}{Koo
  et~al\mbox{.}}{2015}]%
        {koo2015did}
\bibfield{author}{\bibinfo{person}{Jeamin Koo}, \bibinfo{person}{Jungsuk Kwac},
  \bibinfo{person}{Wendy Ju}, \bibinfo{person}{Martin Steinert},
  \bibinfo{person}{Larry Leifer}, {and} \bibinfo{person}{Clifford Nass}.}
  \bibinfo{year}{2015}\natexlab{}.
\newblock \showarticletitle{Why did my car just do that? Explaining
  semi-autonomous driving actions to improve driver understanding, trust, and
  performance}.
\newblock \bibinfo{journal}{\emph{International Journal on Interactive Design
  and Manufacturing (IJIDeM)}} \bibinfo{volume}{9}, \bibinfo{number}{4}
  (\bibinfo{year}{2015}), \bibinfo{pages}{269--275}.
\newblock


\bibitem[\protect\citeauthoryear{Kumar, Paccagnella, Murley, Hennenfent, Mason,
  Bates, and Bailey}{Kumar et~al\mbox{.}}{2018}]%
        {kumar2018skill}
\bibfield{author}{\bibinfo{person}{Deepak Kumar}, \bibinfo{person}{Riccardo
  Paccagnella}, \bibinfo{person}{Paul Murley}, \bibinfo{person}{Eric
  Hennenfent}, \bibinfo{person}{Joshua Mason}, \bibinfo{person}{Adam Bates},
  {and} \bibinfo{person}{Michael Bailey}.} \bibinfo{year}{2018}\natexlab{}.
\newblock \showarticletitle{Skill Squatting Attacks on Amazon Alexa}. In
  \bibinfo{booktitle}{\emph{27th {USENIX} Security Symposium ({USENIX} Security
  18)}}. \bibinfo{publisher}{{USENIX} Association},
  \bibinfo{address}{Baltimore, MD}, \bibinfo{pages}{33--47}.
\newblock
\showISBNx{978-1-939133-04-5}
\urldef\tempurl%
\url{https://www.usenix.org/conference/usenixsecurity18/presentation/kumar}
\showURL{%
\tempurl}


\bibitem[\protect\citeauthoryear{Landis and Koch}{Landis and Koch}{1977}]%
        {landis1977measurement}
\bibfield{author}{\bibinfo{person}{J~Richard Landis} {and}
  \bibinfo{person}{Gary~G Koch}.} \bibinfo{year}{1977}\natexlab{}.
\newblock \showarticletitle{The measurement of observer agreement for
  categorical data}.
\newblock \bibinfo{journal}{\emph{biometrics}} (\bibinfo{year}{1977}),
  \bibinfo{pages}{159--174}.
\newblock


\bibitem[\protect\citeauthoryear{Lau, Zimmerman, and Schaub}{Lau
  et~al\mbox{.}}{2018}]%
        {10.1145/3274371}
\bibfield{author}{\bibinfo{person}{Josephine Lau}, \bibinfo{person}{Benjamin
  Zimmerman}, {and} \bibinfo{person}{Florian Schaub}.}
  \bibinfo{year}{2018}\natexlab{}.
\newblock \showarticletitle{Alexa, Are You Listening? Privacy Perceptions,
  Concerns and Privacy-Seeking Behaviors with Smart Speakers}.
\newblock \bibinfo{journal}{\emph{Proc. ACM Hum.-Comput. Interact.}}
  \bibinfo{volume}{2}, \bibinfo{number}{CSCW}, Article \bibinfo{articleno}{102}
  (\bibinfo{date}{Nov.} \bibinfo{year}{2018}), \bibinfo{numpages}{31}~pages.
\newblock
\urldef\tempurl%
\url{https://doi.org/10.1145/3274371}
\showDOI{\tempurl}


\bibitem[\protect\citeauthoryear{Lee}{Lee}{2018}]%
        {lee2018amazon}
\bibfield{author}{\bibinfo{person}{Dave Lee}.} \bibinfo{year}{2018}\natexlab{}.
\newblock \bibinfo{booktitle}{\emph{Amazon promises fix for creepy Alexa
  laugh}}.
\newblock
\urldef\tempurl%
\url{https://www.bbc.co.uk/news/technology-43325230}
\showURL{%
\tempurl}


\bibitem[\protect\citeauthoryear{Lee, Kim, and Lee}{Lee et~al\mbox{.}}{2019}]%
        {10.1145/3290607.3312796}
\bibfield{author}{\bibinfo{person}{Sunok Lee}, \bibinfo{person}{Sungbae Kim},
  {and} \bibinfo{person}{Sangsu Lee}.} \bibinfo{year}{2019}\natexlab{}.
\newblock \showarticletitle{"What Does Your Agent Look like?": A Drawing Study
  to Understand Users' Perceived Persona of Conversational Agent}. In
  \bibinfo{booktitle}{\emph{Extended Abstracts of the 2019 CHI Conference on
  Human Factors in Computing Systems}} (Glasgow, Scotland Uk)
  \emph{(\bibinfo{series}{CHI EA '19})}. \bibinfo{publisher}{Association for
  Computing Machinery}, \bibinfo{address}{New York, NY, USA},
  \bibinfo{pages}{1–6}.
\newblock
\showISBNx{9781450359719}
\urldef\tempurl%
\url{https://doi.org/10.1145/3290607.3312796}
\showDOI{\tempurl}


\bibitem[\protect\citeauthoryear{Lentzsch, Shah, Andow, Degeling, Das, and
  Enck}{Lentzsch et~al\mbox{.}}{2021}]%
        {lentzsch2021hey}
\bibfield{author}{\bibinfo{person}{Christopher Lentzsch},
  \bibinfo{person}{Sheel~Jayesh Shah}, \bibinfo{person}{Benjamin Andow},
  \bibinfo{person}{Martin Degeling}, \bibinfo{person}{Anupam Das}, {and}
  \bibinfo{person}{William Enck}.} \bibinfo{year}{2021}\natexlab{}.
\newblock \showarticletitle{Hey Alexa, is this Skill Safe?: Taking a Closer
  Look at the Alexa Skill Ecosystem}. In \bibinfo{booktitle}{\emph{28th Annual
  Network and Distributed System Security Symposium (NDSS 2021). The Internet
  Society}}.
\newblock


\bibitem[\protect\citeauthoryear{Lindley, Coulton, and Cooper}{Lindley
  et~al\mbox{.}}{2017}]%
        {lindley2017internet}
\bibfield{author}{\bibinfo{person}{Joseph Lindley}, \bibinfo{person}{Paul
  Coulton}, {and} \bibinfo{person}{Rachel Cooper}.}
  \bibinfo{year}{2017}\natexlab{}.
\newblock \showarticletitle{Why the internet of things needs object orientated
  ontology}.
\newblock \bibinfo{journal}{\emph{The Design Journal}} \bibinfo{volume}{20},
  \bibinfo{number}{sup1} (\bibinfo{year}{2017}), \bibinfo{pages}{S2846--S2857}.
\newblock


\bibitem[\protect\citeauthoryear{Lopatovska, Rink, Knight, Raines, Cosenza,
  Williams, Sorsche, Hirsch, Li, and Martinez}{Lopatovska
  et~al\mbox{.}}{2018}]%
        {lopatovska2018talk}
\bibfield{author}{\bibinfo{person}{Irene Lopatovska}, \bibinfo{person}{Katrina
  Rink}, \bibinfo{person}{Ian Knight}, \bibinfo{person}{Kieran Raines},
  \bibinfo{person}{Kevin Cosenza}, \bibinfo{person}{Harriet Williams},
  \bibinfo{person}{Perachya Sorsche}, \bibinfo{person}{David Hirsch},
  \bibinfo{person}{Qi Li}, {and} \bibinfo{person}{Adrianna Martinez}.}
  \bibinfo{year}{2018}\natexlab{}.
\newblock \showarticletitle{Talk to me: Exploring user interactions with the
  Amazon Alexa}.
\newblock \bibinfo{journal}{\emph{Journal of Librarianship and Information
  Science}} (\bibinfo{year}{2018}).
\newblock


\bibitem[\protect\citeauthoryear{Lopatovska and Williams}{Lopatovska and
  Williams}{2018}]%
        {10.1145/3176349.3176868}
\bibfield{author}{\bibinfo{person}{Irene Lopatovska} {and}
  \bibinfo{person}{Harriet Williams}.} \bibinfo{year}{2018}\natexlab{}.
\newblock \showarticletitle{Personification of the Amazon Alexa: BFF or a
  Mindless Companion}. In \bibinfo{booktitle}{\emph{Proceedings of the 2018
  Conference on Human Information Interaction \& Retrieval}} (New Brunswick,
  NJ, USA) \emph{(\bibinfo{series}{CHIIR '18})}.
  \bibinfo{publisher}{Association for Computing Machinery},
  \bibinfo{address}{New York, NY, USA}, \bibinfo{pages}{265–268}.
\newblock
\showISBNx{9781450349253}
\urldef\tempurl%
\url{https://doi.org/10.1145/3176349.3176868}
\showDOI{\tempurl}


\bibitem[\protect\citeauthoryear{Madsen and Gregor}{Madsen and Gregor}{2000}]%
        {madsen2000measuring}
\bibfield{author}{\bibinfo{person}{Maria Madsen} {and} \bibinfo{person}{Shirley
  Gregor}.} \bibinfo{year}{2000}\natexlab{}.
\newblock \showarticletitle{Measuring human-computer trust}. In
  \bibinfo{booktitle}{\emph{11th australasian conference on information
  systems}}, Vol.~\bibinfo{volume}{53}. Citeseer, \bibinfo{pages}{6--8}.
\newblock


\bibitem[\protect\citeauthoryear{Malhotra, Kim, and Agarwal}{Malhotra
  et~al\mbox{.}}{2004}]%
        {malhotra2004internet}
\bibfield{author}{\bibinfo{person}{Naresh~K Malhotra}, \bibinfo{person}{Sung~S
  Kim}, {and} \bibinfo{person}{James Agarwal}.}
  \bibinfo{year}{2004}\natexlab{}.
\newblock \showarticletitle{Internet users' information privacy concerns
  (IUIPC): The construct, the scale, and a causal model}.
\newblock \bibinfo{journal}{\emph{Information systems research}}
  \bibinfo{volume}{15}, \bibinfo{number}{4} (\bibinfo{year}{2004}),
  \bibinfo{pages}{336--355}.
\newblock


\bibitem[\protect\citeauthoryear{Malkin, Bernd, Johnson, and Egelman}{Malkin
  et~al\mbox{.}}{2018}]%
        {malkin2018can}
\bibfield{author}{\bibinfo{person}{Nathan Malkin}, \bibinfo{person}{Julia
  Bernd}, \bibinfo{person}{Maritza Johnson}, {and} \bibinfo{person}{Serge
  Egelman}.} \bibinfo{year}{2018}\natexlab{}.
\newblock \showarticletitle{“What Can’t Data Be Used For?” Privacy
  Expectations about Smart TVs in the US}. In
  \bibinfo{booktitle}{\emph{Proceedings of the 3rd European Workshop on Usable
  Security (EuroUSEC), London, UK}}.
\newblock


\bibitem[\protect\citeauthoryear{Marky, Voit, St\"{o}ver, Kunze, Schr\"{o}der,
  and M\"{u}hlh\"{a}user}{Marky et~al\mbox{.}}{2020}]%
        {10.1145/3419249.3420164}
\bibfield{author}{\bibinfo{person}{Karola Marky}, \bibinfo{person}{Alexandra
  Voit}, \bibinfo{person}{Alina St\"{o}ver}, \bibinfo{person}{Kai Kunze},
  \bibinfo{person}{Svenja Schr\"{o}der}, {and} \bibinfo{person}{Max
  M\"{u}hlh\"{a}user}.} \bibinfo{year}{2020}\natexlab{}.
\newblock \showarticletitle{”I Don’t Know How to Protect Myself”:
  Understanding Privacy Perceptions Resulting from the Presence of Bystanders
  in Smart Environments} \emph{(\bibinfo{series}{NordiCHI '20})}.
  \bibinfo{publisher}{Association for Computing Machinery},
  \bibinfo{address}{New York, NY, USA}, Article \bibinfo{articleno}{4},
  \bibinfo{numpages}{11}~pages.
\newblock
\showISBNx{9781450375795}
\urldef\tempurl%
\url{https://doi.org/10.1145/3419249.3420164}
\showDOI{\tempurl}


\bibitem[\protect\citeauthoryear{Mayer, Davis, and Schoorman}{Mayer
  et~al\mbox{.}}{1995}]%
        {mayer1995integrative}
\bibfield{author}{\bibinfo{person}{Roger~C Mayer}, \bibinfo{person}{James~H
  Davis}, {and} \bibinfo{person}{F~David Schoorman}.}
  \bibinfo{year}{1995}\natexlab{}.
\newblock \showarticletitle{An integrative model of organizational trust}.
\newblock \bibinfo{journal}{\emph{Academy of management review}}
  \bibinfo{volume}{20}, \bibinfo{number}{3} (\bibinfo{year}{1995}),
  \bibinfo{pages}{709--734}.
\newblock


\bibitem[\protect\citeauthoryear{Mennicken, Zihler, Juldaschewa, Molnar,
  Aggeler, and Huang}{Mennicken et~al\mbox{.}}{2016}]%
        {10.1145/2971648.2971757}
\bibfield{author}{\bibinfo{person}{Sarah Mennicken}, \bibinfo{person}{Oliver
  Zihler}, \bibinfo{person}{Frida Juldaschewa}, \bibinfo{person}{Veronika
  Molnar}, \bibinfo{person}{David Aggeler}, {and} \bibinfo{person}{Elaine~May
  Huang}.} \bibinfo{year}{2016}\natexlab{}.
\newblock \showarticletitle{"It's like Living with a Friendly Stranger":
  Perceptions of Personality Traits in a Smart Home}. In
  \bibinfo{booktitle}{\emph{Proceedings of the 2016 ACM International Joint
  Conference on Pervasive and Ubiquitous Computing}} (Heidelberg, Germany)
  \emph{(\bibinfo{series}{UbiComp '16})}. \bibinfo{publisher}{Association for
  Computing Machinery}, \bibinfo{address}{New York, NY, USA},
  \bibinfo{pages}{120–131}.
\newblock
\showISBNx{9781450344616}
\urldef\tempurl%
\url{https://doi.org/10.1145/2971648.2971757}
\showDOI{\tempurl}


\bibitem[\protect\citeauthoryear{Mittelstadt, Allo, Taddeo, Wachter, and
  Floridi}{Mittelstadt et~al\mbox{.}}{2016}]%
        {mittelstadt2016ethics}
\bibfield{author}{\bibinfo{person}{Brent~Daniel Mittelstadt},
  \bibinfo{person}{Patrick Allo}, \bibinfo{person}{Mariarosaria Taddeo},
  \bibinfo{person}{Sandra Wachter}, {and} \bibinfo{person}{Luciano Floridi}.}
  \bibinfo{year}{2016}\natexlab{}.
\newblock \showarticletitle{The ethics of algorithms: Mapping the debate}.
\newblock \bibinfo{journal}{\emph{Big Data \& Society}} \bibinfo{volume}{3},
  \bibinfo{number}{2} (\bibinfo{year}{2016}),
  \bibinfo{pages}{2053951716679679}.
\newblock


\bibitem[\protect\citeauthoryear{Moon}{Moon}{2000}]%
        {10.1086/209566}
\bibfield{author}{\bibinfo{person}{Youngme Moon}.}
  \bibinfo{year}{2000}\natexlab{}.
\newblock \showarticletitle{{Intimate Exchanges: Using Computers to Elicit
  Self-Disclosure from Consumers}}.
\newblock \bibinfo{journal}{\emph{Journal of Consumer Research}}
  \bibinfo{volume}{26}, \bibinfo{number}{4} (\bibinfo{date}{03}
  \bibinfo{year}{2000}), \bibinfo{pages}{323--339}.
\newblock
\showISSN{0093-5301}
\urldef\tempurl%
\url{https://doi.org/10.1086/209566}
\showDOI{\tempurl}


\bibitem[\protect\citeauthoryear{Mosca and Such}{Mosca and Such}{2021}]%
        {mosca2021elvira}
\bibfield{author}{\bibinfo{person}{Francesca Mosca} {and} \bibinfo{person}{Jose
  Such}.} \bibinfo{year}{2021}\natexlab{}.
\newblock \showarticletitle{ELVIRA: An explainable agent for value and
  utility-driven multiuser privacy}. In \bibinfo{booktitle}{\emph{International
  Conference on Autonomous Agents and Multiagent Systems (AAMAS)}}.
  \bibinfo{pages}{916--924}.
\newblock


\bibitem[\protect\citeauthoryear{Mosca and Such}{Mosca and Such}{2022}]%
        {mosca2022explainable}
\bibfield{author}{\bibinfo{person}{Francesca Mosca} {and} \bibinfo{person}{Jose
  Such}.} \bibinfo{year}{2022}\natexlab{}.
\newblock \showarticletitle{An explainable assistant for multiuser privacy}.
\newblock \bibinfo{journal}{\emph{Autonomous Agents and Multi-Agent Systems
  (JAAMAS)}} \bibinfo{volume}{36}, \bibinfo{number}{1} (\bibinfo{year}{2022}),
  \bibinfo{pages}{1--45}.
\newblock


\bibitem[\protect\citeauthoryear{Murad and Munteanu}{Murad and
  Munteanu}{2020}]%
        {10.1145/3313831.3376522}
\bibfield{author}{\bibinfo{person}{Christine Murad} {and}
  \bibinfo{person}{Cosmin Munteanu}.} \bibinfo{year}{2020}\natexlab{}.
\newblock \bibinfo{booktitle}{\emph{Designing Voice Interfaces: Back to the
  (Curriculum) Basics}}.
\newblock \bibinfo{publisher}{Association for Computing Machinery},
  \bibinfo{address}{New York, NY, USA}, \bibinfo{pages}{1–12}.
\newblock
\showISBNx{9781450367080}
\urldef\tempurl%
\url{https://doi.org/10.1145/3313831.3376522}
\showURL{%
\tempurl}


\bibitem[\protect\citeauthoryear{Naeini, Bhagavatula, Habib, Degeling, Bauer,
  Cranor, and Sadeh}{Naeini et~al\mbox{.}}{2017}]%
        {naeini2017privacy}
\bibfield{author}{\bibinfo{person}{Pardis~Emami Naeini}, \bibinfo{person}{Sruti
  Bhagavatula}, \bibinfo{person}{Hana Habib}, \bibinfo{person}{Martin
  Degeling}, \bibinfo{person}{Lujo Bauer}, \bibinfo{person}{Lorrie~Faith
  Cranor}, {and} \bibinfo{person}{Norman Sadeh}.}
  \bibinfo{year}{2017}\natexlab{}.
\newblock \showarticletitle{Privacy expectations and preferences in an
  $\{$IoT$\}$ world}. In \bibinfo{booktitle}{\emph{Thirteenth Symposium on
  Usable Privacy and Security (SOUPS 2017)}}. \bibinfo{pages}{399--412}.
\newblock


\bibitem[\protect\citeauthoryear{Nass, Steuer, and Tauber}{Nass
  et~al\mbox{.}}{1994}]%
        {10.1145/191666.191703}
\bibfield{author}{\bibinfo{person}{Clifford Nass}, \bibinfo{person}{Jonathan
  Steuer}, {and} \bibinfo{person}{Ellen~R. Tauber}.}
  \bibinfo{year}{1994}\natexlab{}.
\newblock \showarticletitle{Computers Are Social Actors}. In
  \bibinfo{booktitle}{\emph{Proceedings of the SIGCHI Conference on Human
  Factors in Computing Systems}} (Boston, Massachusetts, USA)
  \emph{(\bibinfo{series}{CHI '94})}. \bibinfo{publisher}{Association for
  Computing Machinery}, \bibinfo{address}{New York, NY, USA},
  \bibinfo{pages}{72–78}.
\newblock
\showISBNx{0897916506}
\urldef\tempurl%
\url{https://doi.org/10.1145/191666.191703}
\showDOI{\tempurl}


\bibitem[\protect\citeauthoryear{Nilsson, Crabtree, Fischer, and
  Koleva}{Nilsson et~al\mbox{.}}{2019}]%
        {nilsson2019breaching}
\bibfield{author}{\bibinfo{person}{Tommy Nilsson}, \bibinfo{person}{Andy
  Crabtree}, \bibinfo{person}{Joel Fischer}, {and} \bibinfo{person}{Boriana
  Koleva}.} \bibinfo{year}{2019}\natexlab{}.
\newblock \showarticletitle{Breaching the future: understanding human
  challenges of autonomous systems for the home}.
\newblock \bibinfo{journal}{\emph{Personal and Ubiquitous Computing}}
  \bibinfo{volume}{23}, \bibinfo{number}{2} (\bibinfo{year}{2019}),
  \bibinfo{pages}{287--307}.
\newblock


\bibitem[\protect\citeauthoryear{Oulasvirta, Pihlajamaa, Perki\"{o}, Ray,
  V\"{a}h\"{a}kangas, Hasu, Vainio, and Myllym\"{a}ki}{Oulasvirta
  et~al\mbox{.}}{2012}]%
        {10.1145/2370216.2370224}
\bibfield{author}{\bibinfo{person}{Antti Oulasvirta}, \bibinfo{person}{Aurora
  Pihlajamaa}, \bibinfo{person}{Jukka Perki\"{o}}, \bibinfo{person}{Debarshi
  Ray}, \bibinfo{person}{Taneli V\"{a}h\"{a}kangas}, \bibinfo{person}{Tero
  Hasu}, \bibinfo{person}{Niklas Vainio}, {and} \bibinfo{person}{Petri
  Myllym\"{a}ki}.} \bibinfo{year}{2012}\natexlab{}.
\newblock \showarticletitle{Long-Term Effects of Ubiquitous Surveillance in the
  Home}. In \bibinfo{booktitle}{\emph{Proceedings of the 2012 ACM Conference on
  Ubiquitous Computing}} (Pittsburgh, Pennsylvania)
  \emph{(\bibinfo{series}{UbiComp '12})}. \bibinfo{publisher}{Association for
  Computing Machinery}, \bibinfo{address}{New York, NY, USA},
  \bibinfo{pages}{41–50}.
\newblock
\showISBNx{9781450312240}
\urldef\tempurl%
\url{https://doi.org/10.1145/2370216.2370224}
\showDOI{\tempurl}


\bibitem[\protect\citeauthoryear{Pavlou and Fygenson}{Pavlou and
  Fygenson}{2006}]%
        {pavlou2006understanding}
\bibfield{author}{\bibinfo{person}{Paul~A Pavlou} {and} \bibinfo{person}{Mendel
  Fygenson}.} \bibinfo{year}{2006}\natexlab{}.
\newblock \showarticletitle{Understanding and predicting electronic commerce
  adoption: An extension of the theory of planned behavior}.
\newblock \bibinfo{journal}{\emph{MIS quarterly}} (\bibinfo{year}{2006}),
  \bibinfo{pages}{115--143}.
\newblock


\bibitem[\protect\citeauthoryear{Payr}{Payr}{2013}]%
        {payr2013virtual}
\bibfield{author}{\bibinfo{person}{Sabine Payr}.}
  \bibinfo{year}{2013}\natexlab{}.
\newblock \showarticletitle{Virtual butlers and real people: Styles and
  practices in long-term use of a companion}.
\newblock In \bibinfo{booktitle}{\emph{Your Virtual Butler}}.
  \bibinfo{publisher}{Springer}, \bibinfo{pages}{134--178}.
\newblock


\bibitem[\protect\citeauthoryear{Pierce and DiSalvo}{Pierce and
  DiSalvo}{2018}]%
        {10.1145/3173574.3174123}
\bibfield{author}{\bibinfo{person}{James Pierce} {and} \bibinfo{person}{Carl
  DiSalvo}.} \bibinfo{year}{2018}\natexlab{}.
\newblock \bibinfo{booktitle}{\emph{Addressing Network Anxieties with
  Alternative Design Metaphors}}.
\newblock \bibinfo{publisher}{Association for Computing Machinery},
  \bibinfo{address}{New York, NY, USA}, \bibinfo{pages}{1–13}.
\newblock
\showISBNx{9781450356206}
\urldef\tempurl%
\url{https://doi.org/10.1145/3173574.3174123}
\showURL{%
\tempurl}


\bibitem[\protect\citeauthoryear{Porcheron, Fischer, Reeves, and
  Sharples}{Porcheron et~al\mbox{.}}{2018}]%
        {10.1145/3173574.3174214}
\bibfield{author}{\bibinfo{person}{Martin Porcheron}, \bibinfo{person}{Joel~E.
  Fischer}, \bibinfo{person}{Stuart Reeves}, {and} \bibinfo{person}{Sarah
  Sharples}.} \bibinfo{year}{2018}\natexlab{}.
\newblock \bibinfo{booktitle}{\emph{Voice Interfaces in Everyday Life}}.
\newblock \bibinfo{publisher}{Association for Computing Machinery},
  \bibinfo{address}{New York, NY, USA}, \bibinfo{pages}{1–12}.
\newblock
\showISBNx{9781450356206}
\urldef\tempurl%
\url{https://doi.org/10.1145/3173574.3174214}
\showURL{%
\tempurl}


\bibitem[\protect\citeauthoryear{Pradhan, Findlater, and Lazar}{Pradhan
  et~al\mbox{.}}{2019}]%
        {10.1145/3359316}
\bibfield{author}{\bibinfo{person}{Alisha Pradhan}, \bibinfo{person}{Leah
  Findlater}, {and} \bibinfo{person}{Amanda Lazar}.}
  \bibinfo{year}{2019}\natexlab{}.
\newblock \showarticletitle{"Phantom Friend" or "Just a Box with Information":
  Personification and Ontological Categorization of Smart Speaker-Based Voice
  Assistants by Older Adults}.
\newblock \bibinfo{journal}{\emph{Proc. ACM Hum.-Comput. Interact.}}
  \bibinfo{volume}{3}, \bibinfo{number}{CSCW}, Article \bibinfo{articleno}{214}
  (\bibinfo{date}{Nov.} \bibinfo{year}{2019}), \bibinfo{numpages}{21}~pages.
\newblock
\urldef\tempurl%
\url{https://doi.org/10.1145/3359316}
\showDOI{\tempurl}


\bibitem[\protect\citeauthoryear{Pybus and Cot{\'e}}{Pybus and
  Cot{\'e}}{2021}]%
        {pybus2021did}
\bibfield{author}{\bibinfo{person}{Jennifer Pybus} {and} \bibinfo{person}{M
  Cot{\'e}}.} \bibinfo{year}{2021}\natexlab{}.
\newblock \showarticletitle{Did you give permission? Datafication in the mobile
  ecosystem}.
\newblock \bibinfo{journal}{\emph{Information, Communication \& Society}}
  (\bibinfo{year}{2021}), \bibinfo{pages}{1--19}.
\newblock


\bibitem[\protect\citeauthoryear{Ribeiro, Singh, and Guestrin}{Ribeiro
  et~al\mbox{.}}{2016}]%
        {ribeiro2016should}
\bibfield{author}{\bibinfo{person}{Marco~Tulio Ribeiro},
  \bibinfo{person}{Sameer Singh}, {and} \bibinfo{person}{Carlos Guestrin}.}
  \bibinfo{year}{2016}\natexlab{}.
\newblock \showarticletitle{Why Should {I} Trust You?: Explaining the
  Predictions of Any Classifier}. In \bibinfo{booktitle}{\emph{Proceedings of
  the 22nd ACM SIGKDD International Conference on Knowledge Discovery and Data
  Mining}}. ACM, \bibinfo{pages}{1135--1144}.
\newblock
\urldef\tempurl%
\url{https://doi.org/10.1145/2939672.2939778}
\showDOI{\tempurl}


\bibitem[\protect\citeauthoryear{Sannon, Stoll, DiFranzo, Jung, and
  Bazarova}{Sannon et~al\mbox{.}}{2020}]%
        {10.1145/3375188}
\bibfield{author}{\bibinfo{person}{Shruti Sannon}, \bibinfo{person}{Brett
  Stoll}, \bibinfo{person}{Dominic DiFranzo}, \bibinfo{person}{Malte~F. Jung},
  {and} \bibinfo{person}{Natalya~N. Bazarova}.}
  \bibinfo{year}{2020}\natexlab{}.
\newblock \showarticletitle{“I Just Shared Your Responses”: Extending
  Communication Privacy Management Theory to Interactions with Conversational
  Agents}.
\newblock \bibinfo{journal}{\emph{Proc. ACM Hum.-Comput. Interact.}}
  \bibinfo{volume}{4}, \bibinfo{number}{GROUP}, Article \bibinfo{articleno}{08}
  (\bibinfo{date}{Jan.} \bibinfo{year}{2020}), \bibinfo{numpages}{18}~pages.
\newblock
\urldef\tempurl%
\url{https://doi.org/10.1145/3375188}
\showDOI{\tempurl}


\bibitem[\protect\citeauthoryear{Schaub, Balebako, and Cranor}{Schaub
  et~al\mbox{.}}{2017}]%
        {schaub2017designing}
\bibfield{author}{\bibinfo{person}{Florian Schaub}, \bibinfo{person}{Rebecca
  Balebako}, {and} \bibinfo{person}{Lorrie~Faith Cranor}.}
  \bibinfo{year}{2017}\natexlab{}.
\newblock \showarticletitle{Designing effective privacy notices and controls}.
\newblock \bibinfo{journal}{\emph{IEEE Internet Computing}}
  (\bibinfo{year}{2017}).
\newblock


\bibitem[\protect\citeauthoryear{Sciuto, Saini, Forlizzi, and Hong}{Sciuto
  et~al\mbox{.}}{2018}]%
        {10.1145/3196709.3196772}
\bibfield{author}{\bibinfo{person}{Alex Sciuto}, \bibinfo{person}{Arnita
  Saini}, \bibinfo{person}{Jodi Forlizzi}, {and} \bibinfo{person}{Jason~I.
  Hong}.} \bibinfo{year}{2018}\natexlab{}.
\newblock \showarticletitle{"Hey Alexa, What's Up?": A Mixed-Methods Studies of
  In-Home Conversational Agent Usage}. In \bibinfo{booktitle}{\emph{Proceedings
  of the 2018 Designing Interactive Systems Conference}} (Hong Kong, China)
  \emph{(\bibinfo{series}{DIS '18})}. \bibinfo{publisher}{Association for
  Computing Machinery}, \bibinfo{address}{New York, NY, USA},
  \bibinfo{pages}{857–868}.
\newblock
\showISBNx{9781450351980}
\urldef\tempurl%
\url{https://doi.org/10.1145/3196709.3196772}
\showDOI{\tempurl}


\bibitem[\protect\citeauthoryear{Seymour, Binns, Slovak, Van~Kleek, and
  Shadbolt}{Seymour et~al\mbox{.}}{2020a}]%
        {10.1145/3357236.3395501}
\bibfield{author}{\bibinfo{person}{William Seymour}, \bibinfo{person}{Reuben
  Binns}, \bibinfo{person}{Petr Slovak}, \bibinfo{person}{Max Van~Kleek}, {and}
  \bibinfo{person}{Nigel Shadbolt}.} \bibinfo{year}{2020}\natexlab{a}.
\newblock \showarticletitle{Strangers in the Room: Unpacking Perceptions of
  'Smartness' and Related Ethical Concerns in the Home}. In
  \bibinfo{booktitle}{\emph{Proceedings of the 2020 ACM Designing Interactive
  Systems Conference}} (Eindhoven, Netherlands) \emph{(\bibinfo{series}{DIS
  '20})}. \bibinfo{publisher}{Association for Computing Machinery},
  \bibinfo{address}{New York, NY, USA}, \bibinfo{pages}{841–854}.
\newblock
\showISBNx{9781450369749}
\urldef\tempurl%
\url{https://doi.org/10.1145/3357236.3395501}
\showDOI{\tempurl}


\bibitem[\protect\citeauthoryear{Seymour, Kraemer, Binns, and
  Van~Kleek}{Seymour et~al\mbox{.}}{2020b}]%
        {10.1145/3313831.3376264}
\bibfield{author}{\bibinfo{person}{William Seymour}, \bibinfo{person}{Martin~J.
  Kraemer}, \bibinfo{person}{Reuben Binns}, {and} \bibinfo{person}{Max
  Van~Kleek}.} \bibinfo{year}{2020}\natexlab{b}.
\newblock \showarticletitle{Informing the Design of Privacy-Empowering Tools
  for the Connected Home}. In \bibinfo{booktitle}{\emph{Proceedings of the 2020
  CHI Conference on Human Factors in Computing Systems}} (Honolulu, HI, USA)
  \emph{(\bibinfo{series}{CHI '20})}. \bibinfo{publisher}{Association for
  Computing Machinery}, \bibinfo{address}{New York, NY, USA},
  \bibinfo{pages}{1–14}.
\newblock
\showISBNx{9781450367080}
\urldef\tempurl%
\url{https://doi.org/10.1145/3313831.3376264}
\showDOI{\tempurl}


\bibitem[\protect\citeauthoryear{Seymour and Van~Kleek}{Seymour and
  Van~Kleek}{2021}]%
        {10.1145/3479515}
\bibfield{author}{\bibinfo{person}{William Seymour} {and} \bibinfo{person}{Max
  Van~Kleek}.} \bibinfo{year}{2021}\natexlab{}.
\newblock \showarticletitle{Exploring Interactions Between Trust,
  Anthropomorphism, and Relationship Development in Voice Assistants}.
\newblock \bibinfo{journal}{\emph{Proc. ACM Hum.-Comput. Interact.}}
  \bibinfo{volume}{5}, \bibinfo{number}{CSCW2}, Article
  \bibinfo{articleno}{371} (\bibinfo{date}{Oct.} \bibinfo{year}{2021}),
  \bibinfo{numpages}{16}~pages.
\newblock
\urldef\tempurl%
\url{https://doi.org/10.1145/3479515}
\showDOI{\tempurl}


\bibitem[\protect\citeauthoryear{Shklovski, Mainwaring, Sk\'{u}lad\'{o}ttir,
  and Borgthorsson}{Shklovski et~al\mbox{.}}{2014}]%
        {10.1145/2556288.2557421}
\bibfield{author}{\bibinfo{person}{Irina Shklovski}, \bibinfo{person}{Scott~D.
  Mainwaring}, \bibinfo{person}{Halla~Hrund Sk\'{u}lad\'{o}ttir}, {and}
  \bibinfo{person}{H\"{o}skuldur Borgthorsson}.}
  \bibinfo{year}{2014}\natexlab{}.
\newblock \showarticletitle{Leakiness and Creepiness in App Space: Perceptions
  of Privacy and Mobile App Use}. In \bibinfo{booktitle}{\emph{Proceedings of
  the SIGCHI Conference on Human Factors in Computing Systems}} (Toronto,
  Ontario, Canada) \emph{(\bibinfo{series}{CHI '14})}.
  \bibinfo{publisher}{Association for Computing Machinery},
  \bibinfo{address}{New York, NY, USA}, \bibinfo{pages}{2347–2356}.
\newblock
\showISBNx{9781450324731}
\urldef\tempurl%
\url{https://doi.org/10.1145/2556288.2557421}
\showDOI{\tempurl}


\bibitem[\protect\citeauthoryear{Storer, Judge, and Branham}{Storer
  et~al\mbox{.}}{2020}]%
        {10.1145/3313831.3376225}
\bibfield{author}{\bibinfo{person}{Kevin~M. Storer},
  \bibinfo{person}{Tejinder~K. Judge}, {and} \bibinfo{person}{Stacy~M.
  Branham}.} \bibinfo{year}{2020}\natexlab{}.
\newblock \bibinfo{booktitle}{\emph{"All in the Same Boat": Tradeoffs of Voice
  Assistant Ownership for Mixed-Visual-Ability Families}}.
\newblock \bibinfo{publisher}{Association for Computing Machinery},
  \bibinfo{address}{New York, NY, USA}, \bibinfo{pages}{1–14}.
\newblock
\showISBNx{9781450367080}
\urldef\tempurl%
\url{https://doi.org/10.1145/3313831.3376225}
\showURL{%
\tempurl}


\bibitem[\protect\citeauthoryear{Such}{Such}{2017}]%
        {such2017privacy}
\bibfield{author}{\bibinfo{person}{Jose Such}.}
  \bibinfo{year}{2017}\natexlab{}.
\newblock \showarticletitle{Privacy and autonomous systems}. In
  \bibinfo{booktitle}{\emph{Proceedings of the Twenty-Sixth International Joint
  Conference on Artificial Intelligence, IJCAI-17}}.
  \bibinfo{pages}{4761--4767}.
\newblock


\bibitem[\protect\citeauthoryear{Sugawara, Cyr, Rampazzi, Genkin, and
  Fu}{Sugawara et~al\mbox{.}}{2020}]%
        {sugawara2020light}
\bibfield{author}{\bibinfo{person}{Takeshi Sugawara}, \bibinfo{person}{Benjamin
  Cyr}, \bibinfo{person}{Sara Rampazzi}, \bibinfo{person}{Daniel Genkin}, {and}
  \bibinfo{person}{Kevin Fu}.} \bibinfo{year}{2020}\natexlab{}.
\newblock \showarticletitle{Light commands: laser-based audio injection attacks
  on voice-controllable systems}. In \bibinfo{booktitle}{\emph{29th
  $\{$USENIX$\}$ Security Symposium ($\{$USENIX$\}$ Security 20)}}.
  \bibinfo{pages}{2631--2648}.
\newblock


\bibitem[\protect\citeauthoryear{Van~Kleek, Binns, Zhao, Slack, Lee, Ottewell,
  and Shadbolt}{Van~Kleek et~al\mbox{.}}{2018}]%
        {10.1145/3173574.3173967}
\bibfield{author}{\bibinfo{person}{Max Van~Kleek}, \bibinfo{person}{Reuben
  Binns}, \bibinfo{person}{Jun Zhao}, \bibinfo{person}{Adam Slack},
  \bibinfo{person}{Sauyon Lee}, \bibinfo{person}{Dean Ottewell}, {and}
  \bibinfo{person}{Nigel Shadbolt}.} \bibinfo{year}{2018}\natexlab{}.
\newblock \bibinfo{booktitle}{\emph{X-Ray Refine: Supporting the Exploration
  and Refinement of Information Exposure Resulting from Smartphone Apps}}.
\newblock \bibinfo{publisher}{Association for Computing Machinery},
  \bibinfo{address}{New York, NY, USA}, \bibinfo{pages}{1–13}.
\newblock
\showISBNx{9781450356206}
\urldef\tempurl%
\url{https://doi.org/10.1145/3173574.3173967}
\showURL{%
\tempurl}


\bibitem[\protect\citeauthoryear{Velicer}{Velicer}{1976}]%
        {velicer1976determining}
\bibfield{author}{\bibinfo{person}{Wayne~F Velicer}.}
  \bibinfo{year}{1976}\natexlab{}.
\newblock \showarticletitle{Determining the number of components from the
  matrix of partial correlations}.
\newblock \bibinfo{journal}{\emph{Psychometrika}} \bibinfo{volume}{41},
  \bibinfo{number}{3} (\bibinfo{year}{1976}), \bibinfo{pages}{321--327}.
\newblock


\bibitem[\protect\citeauthoryear{Wachter, Mittelstadt, and Russell}{Wachter
  et~al\mbox{.}}{2017}]%
        {wachter2017counterfactual}
\bibfield{author}{\bibinfo{person}{Sandra Wachter}, \bibinfo{person}{Brent
  Mittelstadt}, {and} \bibinfo{person}{Chris Russell}.}
  \bibinfo{year}{2017}\natexlab{}.
\newblock \showarticletitle{Counterfactual explanations without opening the
  black box: Automated decisions and the GDPR}.
\newblock \bibinfo{journal}{\emph{Harv. JL \& Tech.}}  \bibinfo{volume}{31}
  (\bibinfo{year}{2017}), \bibinfo{pages}{841}.
\newblock


\bibitem[\protect\citeauthoryear{Wirfs-Brock, Mennicken, and Thom}{Wirfs-Brock
  et~al\mbox{.}}{2020}]%
        {10.1145/3313831.3376493}
\bibfield{author}{\bibinfo{person}{Jordan Wirfs-Brock}, \bibinfo{person}{Sarah
  Mennicken}, {and} \bibinfo{person}{Jennifer Thom}.}
  \bibinfo{year}{2020}\natexlab{}.
\newblock \bibinfo{booktitle}{\emph{Giving Voice to Silent Data: Designing with
  Personal Music Listening History}}.
\newblock \bibinfo{publisher}{Association for Computing Machinery},
  \bibinfo{address}{New York, NY, USA}, \bibinfo{pages}{1–11}.
\newblock
\showISBNx{9781450367080}
\urldef\tempurl%
\url{https://doi.org/10.1145/3313831.3376493}
\showURL{%
\tempurl}


\bibitem[\protect\citeauthoryear{Xu and Warschauer}{Xu and Warschauer}{2020}]%
        {10.1145/3313831.3376416}
\bibfield{author}{\bibinfo{person}{Ying Xu} {and} \bibinfo{person}{Mark
  Warschauer}.} \bibinfo{year}{2020}\natexlab{}.
\newblock \showarticletitle{What Are You Talking To?: Understanding Children's
  Perceptions of Conversational Agents}. In
  \bibinfo{booktitle}{\emph{Proceedings of the 2020 CHI Conference on Human
  Factors in Computing Systems}} (Honolulu, HI, USA)
  \emph{(\bibinfo{series}{CHI '20})}. \bibinfo{publisher}{Association for
  Computing Machinery}, \bibinfo{address}{New York, NY, USA},
  \bibinfo{pages}{1–13}.
\newblock
\showISBNx{9781450367080}
\urldef\tempurl%
\url{https://doi.org/10.1145/3313831.3376416}
\showDOI{\tempurl}


\bibitem[\protect\citeauthoryear{Yao, Basdeo, Mcdonough, and Wang}{Yao
  et~al\mbox{.}}{2019}]%
        {10.1145/3359161}
\bibfield{author}{\bibinfo{person}{Yaxing Yao}, \bibinfo{person}{Justin~Reed
  Basdeo}, \bibinfo{person}{Oriana~Rosata Mcdonough}, {and}
  \bibinfo{person}{Yang Wang}.} \bibinfo{year}{2019}\natexlab{}.
\newblock \showarticletitle{Privacy Perceptions and Designs of Bystanders in
  Smart Homes}.
\newblock  \bibinfo{volume}{3}, \bibinfo{number}{CSCW}, Article
  \bibinfo{articleno}{59} (\bibinfo{date}{Nov.} \bibinfo{year}{2019}),
  \bibinfo{numpages}{24}~pages.
\newblock
\urldef\tempurl%
\url{https://doi.org/10.1145/3359161}
\showDOI{\tempurl}


\bibitem[\protect\citeauthoryear{Zeng, Mare, and Roesner}{Zeng
  et~al\mbox{.}}{2017}]%
        {zeng2017end}
\bibfield{author}{\bibinfo{person}{Eric Zeng}, \bibinfo{person}{Shrirang Mare},
  {and} \bibinfo{person}{Franziska Roesner}.} \bibinfo{year}{2017}\natexlab{}.
\newblock \showarticletitle{End user security and privacy concerns with smart
  homes}. In \bibinfo{booktitle}{\emph{thirteenth symposium on usable privacy
  and security ($\{$SOUPS$\}$ 2017)}}. \bibinfo{pages}{65--80}.
\newblock


\bibitem[\protect\citeauthoryear{Zhan, Sarkadi, Criado, and Such}{Zhan
  et~al\mbox{.}}{2022}]%
        {zhan2022model}
\bibfield{author}{\bibinfo{person}{Xiao Zhan}, \bibinfo{person}{Stefan
  Sarkadi}, \bibinfo{person}{Natalia Criado}, {and} \bibinfo{person}{Jose
  Such}.} \bibinfo{year}{2022}\natexlab{}.
\newblock \showarticletitle{A Model for Governing Information Sharing in Smart
  Assistants}. In \bibinfo{booktitle}{\emph{AAAI/ACM Conference on AI, Ethics,
  and Society (AIES)}}.
\newblock


\bibitem[\protect\citeauthoryear{Zhang, Yan, Ji, Zhang, Zhang, and Xu}{Zhang
  et~al\mbox{.}}{2017}]%
        {10.1145/3133956.3134052}
\bibfield{author}{\bibinfo{person}{Guoming Zhang}, \bibinfo{person}{Chen Yan},
  \bibinfo{person}{Xiaoyu Ji}, \bibinfo{person}{Tianchen Zhang},
  \bibinfo{person}{Taimin Zhang}, {and} \bibinfo{person}{Wenyuan Xu}.}
  \bibinfo{year}{2017}\natexlab{}.
\newblock \bibinfo{booktitle}{\emph{DolphinAttack: Inaudible Voice Commands}}.
\newblock \bibinfo{publisher}{Association for Computing Machinery},
  \bibinfo{address}{New York, NY, USA}, \bibinfo{pages}{103–117}.
\newblock
\showISBNx{9781450349468}
\urldef\tempurl%
\url{https://doi.org/10.1145/3133956.3134052}
\showURL{%
\tempurl}


\bibitem[\protect\citeauthoryear{Zimmermann, Gerber, Marky, B{\"o}ck, and
  Kirchbuchner}{Zimmermann et~al\mbox{.}}{2019}]%
        {zimmermann2019assessing}
\bibfield{author}{\bibinfo{person}{Verena Zimmermann}, \bibinfo{person}{Paul
  Gerber}, \bibinfo{person}{Karola Marky}, \bibinfo{person}{Leon B{\"o}ck},
  {and} \bibinfo{person}{Florian Kirchbuchner}.}
  \bibinfo{year}{2019}\natexlab{}.
\newblock \showarticletitle{Assessing users’ privacy and security concerns of
  smart home technologies}.
\newblock \bibinfo{journal}{\emph{i-com}} \bibinfo{volume}{18},
  \bibinfo{number}{3} (\bibinfo{year}{2019}), \bibinfo{pages}{197--216}.
\newblock


\end{thebibliography}

\end{document}